\documentclass[12pt]{article}

\usepackage{latexsym}
\usepackage{amssymb}
\usepackage{amsbsy}
\usepackage{amsmath}
\usepackage{cite}
\usepackage{epsfig}
\usepackage{graphics,color}
\usepackage{a4}
\numberwithin{equation}{section}

\allowdisplaybreaks

\newcommand{\as}{\alpha_s}
\newcommand{\hm}{\hspace*{-0.6cm}}

\newcommand{\be}{\begin{equation}}
\newcommand{\ee}{\end{equation}}

\newcommand{\bea}{\begin{eqnarray}}
\newcommand{\eea}{\end{eqnarray}}
\newcommand{\bean}{\begin{eqnarray*}}
\newcommand{\eean}{\end{eqnarray*}}

\newcommand{\om} {\omega}

\newcommand{\pv}{{\mathbf p}}
\newcommand{\xv}{{\mathbf x}}
\newcommand{\vecnul}{{\mathbf 0}}

\newcommand{\nn}{\nonumber}

\newcommand{\bB}{{\bf B}}
\newcommand{\bD}{{\bf D}}
\newcommand{\bDl}{{\bf\Delta}}
\newcommand{\bE}{{\bf E}}
\newcommand{\bsigma}{{\boldsymbol\sigma}}

\begin{document}

\title{\bf\large
What happens to the $\Upsilon$ and $\eta_b$ in the quark-gluon plasma?
Bottomonium spectral functions from lattice QCD
}

\author{G.~Aarts${}^a$, C.~Allton${}^a$, S.~Kim${}^{b,a}$,  M.~P.~Lombardo${}^{c,d}$, M.~B.~Oktay${}^e$, \\
S.~M.~Ryan${}^f$, D.~K.~Sinclair${}^g$ and  J.-I.~Skullerud${}^h$ \\
\mbox{} \\
\mbox{} \\
{${}^a$\em\normalsize Department of Physics, Swansea University, Swansea, United Kingdom} \\
{${}^b$\em\normalsize Department of Physics, Sejong University, Seoul 143-747, Korea} \\
{${}^c$\em\normalsize INFN-Laboratori Nazionali di Frascati, I-00044, Frascati (RM) Italy} \\
{${}^d$\em\normalsize  Humboldt-Universit\"at zu Berlin, 12489 Berlin,Germany} \\
{${}^e$\em\normalsize Physics Department, University of Utah, Salt Lake City,  Utah, USA} \\
{${}^f$\em\normalsize School of Mathematics, Trinity College, Dublin 2, Ireland} \\
{${}^g$\em\normalsize HEP Division, Argonne National Laboratory, 9700 South Cass Avenue} \\
{\em\normalsize Argonne, Illinois 60439, USA} \\
{${}^h$\em\normalsize Department of Mathematical Physics, National University of Ireland Maynooth} \\ 
{\em\normalsize Maynooth, County Kildare, Ireland}
}

\date{September 21, 2011}

\maketitle

\begin{abstract}

We study bottomonium spectral functions in the quark-gluon plasma in
the $\Upsilon$ and $\eta_b$ channels, using lattice QCD simulations
with two flavours of light quark on highly anisotropic lattices.  The
bottom quark is treated with nonrelativistic QCD (NRQCD).  In the
temperature range we consider, $0.42\leq T/T_c \leq 2.09$, we find
that the ground states survive, whereas the excited states are
suppressed as the temperature is increased. The position and width of
the ground states are compared to analytical effective field theory
(EFT) predictions. Systematic uncertainties of the maximum entropy
method (MEM), used to construct the spectral functions, are discussed
in some detail.

\end{abstract}


\maketitle


\newpage
 
\section{Introduction}
 \label{sec:intro}

Quarkonia (heavy quark--antiquark bound states) are among the most
important probes of the hot medium created in relativistic heavy ion
collisions.  Unlike light quarks, heavy quarks are predominantly
created in the primordial hard collisions, and do not reach chemical
equilibrium with the medium.  Since $J/\psi$ suppression was proposed
in 1986 as a signature of the formation of the quark-gluon plasma
\cite{Matsui:1986dk}, the charmonium system has been investigated
intensively, both experimentally \cite{Arnaldi:2009ph,Adare:2008sh}
and theoretically \cite{Satz:2006kba,Rapp:2008tf}.  With the advent of
the Large Hadron Collider, there has been increasing interest in
bottomonium states as well, since $b$ quarks are now for the first
time being produced copiously in heavy ion collisions. In particular,
the recent results from CMS indicate the survival of the
$\Upsilon(1S)$ state, but suppression of the $\Upsilon(2S+3S)$ states
\cite{Chatrchyan:2011pe} (see Ref.\ \cite{Reed:2011zz}  for results
from STAR). A number of phenomenological studies have since
attempted to explain this suppression pattern
\cite{Strickland:2011mw,Brezinski:2011ju}. It is generally expected
that bottomonium provides a cleaner probe than charmonium, since
statistical recombination of independent quarks and antiquarks plays a
much less important role, and also since $b$ quarks retain their
nature as `hard probes' to a larger extent.
Cold nuclear matter effects are also expected to be simpler
for $b$ than for $c$ quarks.

Theoretically, quarkonium suppression has traditionally been investigated with potential models (see e.g.\ Refs.\ \cite{Mocsy:2007yj,Mocsy:2007jz} and references therein) and with lattice QCD computations of quarkonium spectral functions \cite{Umeda:2002vr,Asakawa:2003re,Datta:2003ww,Jakovac:2006sf,Aarts:2007pk,Oktay:2010tf,Ding:2010yz,Ohno:2011zc}.  Exploiting the strongly-coupled nature of the quark-gluon plasma,  studies using gauge-gravity duality have also been used, see e.g.\ Refs.\ \cite{Noronha:2009da,Grigoryan:2010pj,Grigoryan:2011cn} for recent results.
Other studies can be found in Refs.\ \cite{Narodetskiy:2011ue,Marasinghe:2011bt}

 In the past few years, the theoretical understanding of quarkonium melting and in-medium modification has been improved substantially by casting the problem in the language of effective field theories (EFTs) 
 \cite{Laine:2006ns,Laine:2007gj,Burnier:2007qm,Laine:2008cf,Laine:2011xr,Brambilla:2008cx,Brambilla:2010vq,Brambilla:2010xn,Brambilla:2011mk,Beraudo:2007ky,Beraudo:2010tw}. By relying on scale separation between the heavy quark mass $M$ and the temperature $T$  of the quark-gluon plasma and on  weak coupling to distinguish the  
inverse system size $M\as$, the binding energy $M\as^2$ and the inverse Debye screening length $m_D\sim \sqrt{\as}T$, a series of EFTs can be written down.
One of the main outcomes of this formulation is the appearance of a complex heavy quark potential, where the imaginary part is generated by integrating out thermal fluctuations. For complex potential model studies, see e.g.\ Refs.\ \cite{Petreczky:2010tk,Margotta:2011ta} as well as those listed above.
Attempts at extracting the complex potential from lattice QCD can be found in Refs.\ \cite{Rothkopf:2009pk,Rothkopf:2011db}.

Various sequences of EFTs can be constructed, depending on the ordering of the scales. For instance, Refs.\
 \cite{Laine:2006ns,Laine:2007gj,Burnier:2007qm} have focused on high temperature, where the bound state is about to fall apart, employing 
\be
 M\gg T> M\as > m_D \gg M\as^2.
  \ee
The corresponding bound state spectral functions are then considerably affected by the presence of the quark-gluon plasma.
On the other hand, in Ref.\ \cite{Brambilla:2010vq}  lower temperatures are considered, using
\be
M\gg  M\as \gg T \gg M\as^2\gg m_D.
\ee
 In this case the ground states are less affected and thermal effects can be cast in terms of  thermal mass shifts and widths.
 
In all cases, integrating out the heavy quark mass scale $M$ yields
nonrelativistic QCD (NRQCD) as an effective field theory.  The
appearance of further EFTs depends on the ordering of the scales and
weak coupling.  Since the applicability of weak coupling arguments is
not guaranteed for temperatures up to $2-3T_c$, where $T_c$ is the
transition temperature between the hadronic phase and the quark-gluon
plasma, it would be desirable to solve NRQCD in the quark-gluon plasma
nonperturbatively, using lattice QCD simulations at finite
temperature. This programme was recently initiated by us.\footnote{See
  Ref.\ \cite{Fingberg:1997qd} for an early pioneering study.}

In Ref.\ \cite{Aarts:2010ek} we studied $S$ and $P$ wave bottomonium
correlators at four different temperatures, $T/T_c=0.42, 1.05, 1.40$
and $2.09$, using NRQCD for the heavy quark dynamics and relativistic
two-flavour lattice simulations for the quark-gluon system.  The main
result of that analysis was the presence of strong temperature
dependence in the $P$ wave correlators (in the $\chi_{b1, b2, b3}$
channels), indicating a melting of $P$ wave bound states in the
quark-gluon plasma. At the highest temperature, we found the behaviour
of the $P$ wave correlators to be consistent with nearly-free dynamics
of the heavy quarks.  

The temperature dependence in the $S$ wave correlators was much less
visible. The goal of this paper is to analyse in detail the $S$ wave
correlators, in the vector ($\Upsilon$) and pseudoscalar ($\eta_b$)
channels, and construct the corresponding spectral functions at
several temperatures in the hadronic phase and the quark-gluon plasma.
 Our main results can be seen in Fig.\ \ref{fig:rho_all_upsilon}, which shows that as the temperature is increased
the ground state peaks of the $\Upsilon$ and $\eta_b$ remain visible,
even though they broaden and reduce in height, while their excited
states become suppressed at higher temperature and are no longer
discernible at $T/T_c\sim 1.68$.  The temperature dependence of the
position and width of the ground state peaks is compared to analytical predictions
obtained within the EFT formalism \cite{Brambilla:2010vq}.  
We note that the survival of the $\Upsilon(1S)$
state and suppression of excited states is consistent with the recent
CMS and STAR results \cite{Chatrchyan:2011pe,Reed:2011zz}.

This paper is organised as follows. In the following section we
decribe NRQCD as an effective field theory for QCD, focusing on finite
temperature aspects. We discuss in particular how the nonrelativistic
formulation has several advantages compared to standard relativistic
dynamics at nonzero temperature.  Lattice details are collected in
Sec.~\ref{sec:lattice}. The main part of the paper starts in
Sec.~\ref{sec:corr}, where the high-precision euclidean correlators in
the $\Upsilon$ and $\eta_b$ channels are presented. The corresponding
spectral functions are shown in Sec.~\ref{sec:spec}. Here we also
compare our results with analytical EFT predictions.  A discussion of
the maximum entropy method \cite{Asakawa:2000tr}, used to construct
the spectral functions, and of systematic uncertainties is given in
Sec.~\ref{sec:sys}. We summarise in Sec.~\ref{sec:conclusion}. The Appendix
contains an analysis of lattice artefacts in NRQCD spectral functions
in the absence of interactions.  Some preliminary results have
previously been presented in Ref.~\cite{Aarts:2011kr}.

\section{NRQCD at nonzero temperature}
\label{sec:NRQCD}

NRQCD is an effective theory of QCD where physics above the scale of
the heavy quark mass is integrated out
\cite{Caswell:1985ui,Lepage:1992tx,Bodwin:1994jh,Brambilla:2004jw}. It
differs from heavy quark effective theory (HQET) in that terms in the
NRQCD lagrangian are ordered in powers of $v = |\pv|/M$, the typical
velocity of a heavy quark in the heavy quarkonium rest
frame. In principle, there are infinitely many terms in such an
expansion and taking into consideration more terms would mean more
accurate relativistic corrections. However, in practice, only a small
number of terms is necessary for a given accuracy, since $v^2$ is
small ($ \sim 0.1$ for the bottom quark) and the series converges
quickly. Also, including more terms in an effective theory would
normally mean tuning more coefficients, resulting in a loss of
predictive power. In practice, the coefficients of the NRQCD
lagrangian can be calculated using perturbation theory since $M$ is
large ($\sim 5$ GeV for the bottom quark) and often tree level values
are sufficient when simulation parameters are chosen judiciously.

In this work, we use the following ${\cal O}(v^4)$ euclidean NRQCD lagrangian
density for the bottom quark \cite{Bodwin:1994jh},
\be
\label{LNRQCD}
{\cal L} = {\cal L}_0 + \delta {\cal L},
\ee
with
\be
\label{LNRQCD_1}
{\cal L}_0 = \psi^\dagger \left(D_\tau - \frac{\bD^2}{2M} \right) \psi +
\chi^\dagger \left(D_\tau + \frac{\bD^2}{2M} \right) \chi,
\ee
and
\begin{eqnarray}
\label{LNRQCD_2}
\delta {\cal L} = &&\hm - \frac{c_1}{8M^3} \left[\psi^\dagger (\bD^2)^2
  \psi - \chi^\dagger (\bD^2)^2 \chi \right] \nonumber \\
&&\hm + c_2 \frac{ig}{8M^2}\left[\psi^\dagger \left(\bD\cdot\bE -
  \bE\cdot\bD\right) \psi + \chi^\dagger \left(\bD\cdot\bE -
  \bE\cdot\bD \right) \chi \right] \nonumber \\
&&\hm - c_3 \frac{g}{8M^2}\left[\psi^\dagger
  \bsigma\cdot\left(\bD\times\bE-\bE\times\bD\right)\psi +
  \chi^\dagger  \bsigma\cdot\left(\bD\times\bE-\bE\times\bD\right)\chi
  \right] \nonumber \\
&&\hm - c_4 \frac{g}{2M} \left[\psi^\dagger \bsigma\cdot\bB \psi -
    \chi^\dagger \bsigma\cdot\bB \chi \right].
\end{eqnarray}
Here $D_\tau$ and $\bD$ are gauge covariant temporal and
spatial derivatives, $\psi$ is the heavy quark and $\chi$ is the
heavy anti-quark.  The coefficients $c_i=1$ at tree level.

In contrast to the relativistic theory, the time evolution for
nonrelativistic heavy quarks is an initial value problem. In
particular, the presence of a nonzero temperature is not imposed as a
boundary condition in the temporal direction of the heavy quark
field. Instead, the effects of temperature enter for the heavy quarks
as they propagate through the thermal medium of light quarks and
gluons. Since we are considering temperatures $T \ll M$, we expect
that finite temperature can be taken into account without affecting
the effective field theory nature of NRQCD.

The absence of thermal boundary conditions  simplifies spectral relations considerably. In the relativistic formulation, the euclidean correlator and its spectral function are related via
 \be
 \label{eq:K}
 G(\tau) = \int_{-\infty}^\infty \frac{d\omega}{2\pi}\,  K(\tau,\om)  \rho(\omega),
 \ee
 with the kernel
 \be
  K(\tau,\om) = \frac{\cosh\left[\omega(\tau-1/2T)\right]}{\sinh\left(\omega/2T\right)}.
 \ee
 Temperature dependence enters in two ways: kinematically due to the
 periodic boundary conditions, reflected in the periodicity of the
 kernel, and dynamically due to the propagation through a
 temperature-dependent medium. It is important to disentangle these
 two, since the first one is present even in the absence of
 interactions and does not reflect the effects of the thermal medium.
 
In NRQCD the kinematical temperature dependence is absent. This can be seen in a number of ways. Following Ref.\ 
 \cite{Burnier:2007qm}, we write $\om=2M+\om'$ and drop terms that 
are exponentially suppressed when $M\gg T$. The spectral relation (\ref{eq:K}) then reduces to
 \be
 G(\tau) =
 \int_{-2M}^\infty\frac{d\omega'}{2\pi}\, \exp(-\om'\tau) \rho(\omega')
\;\;\;\;\; \;\;\;\;\;\;\;\;\;\;(\text{NRQCD}), 
 \ee
 even at nonzero temperature. This reflects  the fact that the NRQCD propagator is constructed from an initial-value problem. Physically it implies that the heavy quarks are not in thermal equilibrium with the light-quark--gluon system, but instead appear as probes.
 
This simplification also removes the problems associated with the
so-called constant contribution \cite{Umeda:2007hy}.  In the small energy limit, the
product of the relativistic kernel and the spectral density is
independent of euclidean time \cite{Aarts:2002cc},
 \begin{equation}
 \lim_{\om\to 0} K(\tau,\om) \rho(\om)= 2T\frac{\rho(\om)}{\om}\Big|_{\om=0},
 \end{equation}
where we used that the spectral function $\rho(\om)$ is an odd function in $\om$ and increases linearly at small $\om$. This is relevant for transport coefficients \cite{Aarts:2002cc}
and for conserved charges, in the presence of which spectral functions
will have a contribution of the form
\begin{equation}
\rho(\om) = \chi2\pi\om\delta(\om) +  \mbox{contribution at larger $\om$},
\end{equation}
where $\chi$ is a susceptibility. 
Spectral weight at vanishing
energy will therefore yield a constant, additive contribution to the
euclidean correlator. It has been argued \cite{Umeda:2007hy,Petreczky:2008px} that this
constant contribution interferes with the interpretation of charmonium
survival or melting, as seen by lattice QCD simulations \cite{Umeda:2002vr,Asakawa:2003re,Datta:2003ww,Jakovac:2006sf,Aarts:2007pk,Oktay:2010tf,Ding:2010yz,Ohno:2011zc}.
It also requires a modification \cite{Aarts:2007wj} of Bryan's algorithm in the implementation of the maximum entropy method \cite{Asakawa:2000tr,Bryan}.
 In NRQCD, the contribution at small energies is absent, since only energies above $2M$ are 
 present.\footnote{Heavy quark diffusion has been studied using heavy quark effective theory \cite{CaronHuot:2009uh,Meyer:2010tt}.}
 In summary, in the heavy quark limit the spectral relation
simplifies considerably, removing the problems associated with thermal
boundary conditions. All temperature effects seen in the
correlators are thus due to changes in the light-quark--gluon system.

\section{Lattice formulation}
\label{sec:lattice}

We solve NRQCD nonpertubatively using lattice QCD, and let the bottom
quarks propagate through a medium of gluons and two flavours of light quark.
Gauge configurations 
with $N_f=2$ dynamical light Wilson-type quark flavours are produced on highly 
anisotropic lattices ($\xi\equiv a_s/a_\tau=6$) of size $N_s^3\times N_\tau$. 
A summary of the lattice datasets is given in Table \ref{tab:latticedetail}, while more 
details of the lattice action and parameters can be found in 
Refs.~\cite{Aarts:2007pk,Morrin:2006tf}. In the light quark sector, $m_\pi/m_\rho\simeq 0.54$, which implies that the light quark masses are comparable to the strange quark mass.

\begin{table}[h]
\begin{center}
\begin{tabular}{ccccr}
\hline\hline 
 \multicolumn{1}{c}{$N_s$} & \multicolumn{1}{c}{$N_\tau$} & \multicolumn{1}{c}{$T$(MeV)} &
\multicolumn{1}{c}{$T/T_c$} & \multicolumn{1}{c}{$N_{\rm cfg}$}
\\ \hline 
12 & 80 &  90  & 0.42 &   250  \\      
12 & 32 & 230 & 1.05 & 1000 \\ 
12 & 28 & 263 & 1.20 & 1000  \\ 
12 & 24 & 306 & 1.40 &    500  \\ 
12 & 20 & 368 & 1.68 & 1000  \\ 
12 & 18 & 408 & 1.86 & 1000  \\ 
12 & 16 & 458 & 2.09 & 1000  \\ 
\hline
\end{tabular}
 \caption{Summary of the lattice data set. The lattice spacing is set 
using the $1P-1S$ spin-averaged splitting in
charmonium~\cite{Oktay:2010tf}, corresponding to 
$a_s\simeq 0.162$fm, $a_\tau^{-1}\simeq 7.35$ GeV. The anisotropy is $a_s/a_\tau=6$.
}  
\label{tab:latticedetail}
\end{center}
\end{table}

 In Ref.\ \cite{Aarts:2010ek} we studied the effect of temperature on $S$ and $P$ wave bottomonium correlators. Compared to that study, we have  greatly increased the number of configurations and expanded the number of temperature values (in Ref.\  \cite{Aarts:2010ek} we only considered $T/T_c=0.42, 1.05, 1.40$ and $2.09$). We have also improved the nonperturbative tuning of the bare anisotropy in the action.\footnote{All data analysed here correspond to Run 7 in the terminology of Ref.\ \cite{Aarts:2007pk}.}

In order for NRQCD to be a consistent effective field theory in a lattice simulation,
the lattice spacing $a_s$, acting as a short-distance cutoff, has to be kept finite and 
satisfy $M a_s \sim 1$. Finite lattice spacing errors can then be
systematically improved as they appear at  the same order as relativistic
effects due to $a_s \sim 1/M$. There are many equivalent
discretizations of the continuum NRQCD lagrangian density discussed above. 
Following earlier studies of heavy quarkonium spectroscopy at zero temperature
\cite{Davies:1994mp,Davies:1995db,Davies:1998im}, we calculate the heavy quark 
Green function on an anisotropic lattice using
\begin{eqnarray}
G (\xv, \tau=0) = &&\hm S(\xv), \nn\\
G (\xv, \tau=a_\tau) = &&\hm  \left(1 - \frac{H_0}{2n}\right)^n
U_4^\dagger(\xv, 0) \left(1 - \frac{H_0}{2n}\right)^n G(\xv,0), \nn \\
G (\xv, \tau+a_\tau) = &&\hm  \left(1 - \frac{H_0}{2n}\right)^n
U_4^\dagger(\xv, \tau) \left(1 - \frac{H_0}{2n}\right)^n 
\left(1 -\delta H \right)  G (\xv, \tau),
\end{eqnarray}
where $S(\xv)$ is the source, the lowest-order hamiltonian reads
\be
H_0 = - \frac{\Delta^{(2)}}{2M},
\ee
and
\begin{eqnarray}
\delta H = &&\hm - \frac{(\Delta^{(2)})^2}{8 M^3} + \frac{ig}{8 M^2}
(\bDl\cdot \bE - \bE\cdot \bDl) - \frac{g}{8 M^2} \bsigma \cdot
  (\bDl\times \bE - \bE\times \bDl)   \nonumber \\
&&\hm
- \frac{g}{2 M} \bsigma\cdot\bB 
 + \frac{a_s^2\Delta^{(4)}}{24 M} - \frac{a_s(\Delta^{(2)})^2}{16 n M^2}.
\label{eq:deltaH}
\end{eqnarray}
The integer $n$ controls the high-momentum behaviour of the
evolution equation. Since the bottom quark is
heavy enough,  we take $n=1$. 
The last two terms in $\delta H$ are corrections to the
kinetic energy term at finite lattice spacing 
\cite{Lepage:1992tx}.  The lattice covariant
derivatives are  defined as
\begin{align}
\Delta_i\psi &= \frac{1}{2a_s}\left[ U_i(x)\psi(x+\hat\imath) -  U_i^\dagger(x-\hat\imath)\psi(x-\hat\imath) \right], \nn \\
 \Delta^{(2)}\psi &= \sum_i \Delta_i^{(2)}
\psi = \sum_i \frac{1}{a_s^2}\left[U_i (x) \psi (x+\hat{\imath}) - 2 \psi (x) +
  U_i^\dagger (x - \hat{\imath}) \psi (x - \hat{\imath})\right],  \nn \\
\Delta^{(4)}\psi &= \sum_i (\Delta_i^{(2)})^2 \psi,
\end{align}
and $\bE$ and $\bB$ in Eq.\ \eqref{eq:deltaH} are lattice definitions of the
chromoelectric and chromomagnetic fields.
We use tadpole improvement \cite{Lepage:1992xa}:
\be 
U_i (x) \rightarrow \frac{U_i (x)}{u_s}, 
\;\;\;\;\;\; 
U_0 (x) \rightarrow \frac{U_0 (x)}{u_\tau}, 
\ee 
where $u_{s,\tau}$ are the average space-like ($s$)
and time-like ($\tau$) links, determined from the plaquette
expectation values; although in practice, the time-like mean link $u_\tau$ is
set to 1. The coefficients $c_i$ are then set to $1$.
Note that   $u_s\neq u_\tau$, since the lattice is anisotropic.

An accurate determination of bottomonium spectroscopy requires careful 
tuning of the bare heavy quark mass $M$ to satisfy NRQCD dispersion 
relations \cite{Davies:1994mp}. To 
study the finite-temperature modification of NRQCD propagators, an approximate 
choice of $a_sM$ such that $M \simeq 5$ GeV is sufficient.

There are many sources of systematic error in a lattice NRQCD
calculation \cite{Davies:1998im}. The three usual ones are effects of finite lattice spacing, finite
volume and light quark vacuum polarization. Two more arise from the 
effective field theory nature: relativistic effects and radiative
corrections to the couplings in the NRQCD expansion.  Among
these, finite volume effects are expected to be small for bottomonium since the
physical size of bottomonium is small. Relativistic effects from the
neglected higher order terms beyond ${\cal O} (v^4)$ are expected to be 
small as well since $v^2 \sim 0.1$. From experience
in heavy quarkonium spectroscopy \cite{Davies:1994mp,Davies:1995db},
tadpole improvement is expected to reduce radiative corrections.
The light quark masses used in this work are 
somewhat large but since the effects of the light quark vacuum polarization in
heavy quarkonium physics is small, systematic effects from this
are expected to be minor. The presence of a finite lattice spacing with
the condition $M a_s \sim 1$ is a well-known issue in lattice NRQCD, ruling out a
continuum limit of NRQCD results, but at finite temperature this problem
is no worse than at zero temperature. In all,
the qualitative features of the finite temperature behaviour of
bottomonium reported in this work are expected to remain valid even
after the careful consideration of systematic errors.

\section{Correlators}
\label{sec:corr}

\begin{figure}[h]
\begin{center}
 \includegraphics*[width=\textwidth]{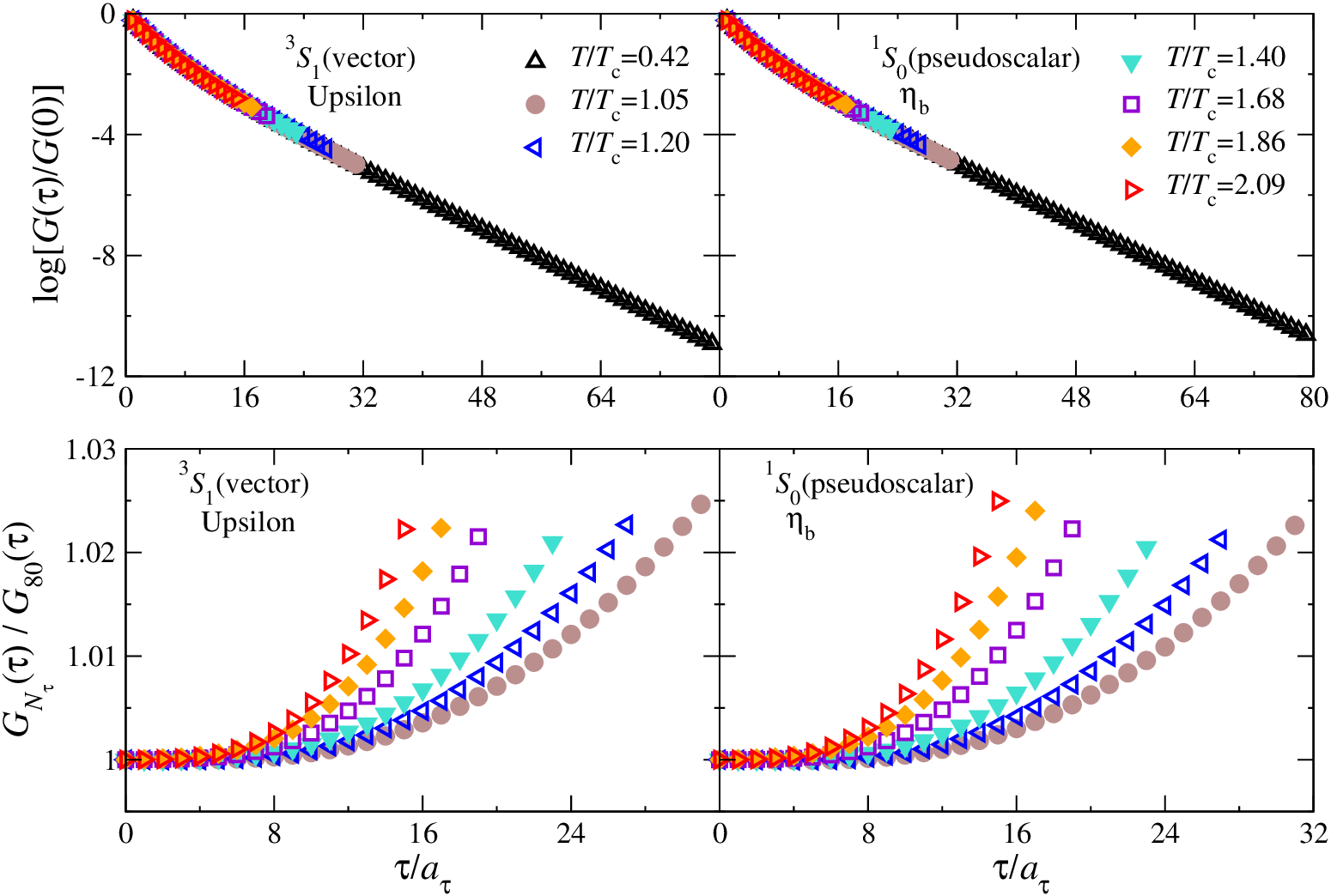}
\end{center}
 \caption{
Euclidean correlation functions $G(\tau)$ as a function of the euclidean time in the vector ($\Upsilon$)  channel (left) and the pseudoscalar ($\eta_b$) channel (right), for all temperatures available, using point sources. At the top the correlators are normalised  with the value at $\tau=0$ and  shown on a logarithmic scale, while on the bottom the high-temperature correlators are normalised with the correlator at the lowest temperature, $T/T_c=0.42$ ($N_\tau=80$). The errors are smaller than the symbols.
}
 \label{fig:G}
 \end{figure}

The starting point for the remainder of this paper is the
high-precision euclidean correlators in the vector ($\Upsilon$) and
the pseudoscalar ($\eta_b$) channel, obtained by solving the NRQCD
evolution equations. In Fig.\ \ref{fig:G} (top) we show these
correlators, normalised with the value at $\tau=0$ and on a
logarithmic scale, for all temperatures available. Due to the use of
NRQCD, there is no periodicity in the temporal direction. Combined
with the large anisotropy, this implies that many temporal lattice
points are available for the analysis, even at the highest
temperature.

It is clear from the plots on the top that the temperature dependence
is very mild. In order to make the temperature dependence visible, we
show on the bottom the ratio of the high-temperature correlators with
the one in the hadronic phase ($T/T_c=0.42$, $N_\tau=80$). We observe
that the effect of increasing the temperature is monotonic and always
below 3\%.  We remark here that the ratios depend on the sources used
in the NRQCD evolution; the results shown here are obtained with point
sources. Since potential model calculations typically use point
(delta function) sources as well, a comparison between potential model
predictions and our NRQCD results should be possible.

In contrast to the relativistic case, the temperature dependence  seen here does not arise from the thermal boundary conditions,  but solely from the presence of the medium of gluons and light quarks at different temperatures. In terms of the spectral relation,
\be
 \label{eq:Gnr}
G(\tau) = \int\frac{d\om}{2\pi}\, e^{-\om\tau}\rho(\om),
\ee
 this is reflected in the  temperature-independent kernel $e^{-\om\tau}$. In the relativistic case, the temperature dependence of the kernel 
means a direct comparison between correlators at different temperatures is not straightforward. Often the kinematical temperature dependence is eliminated by using so-called reconstructed correlators, which requires the calculation of a spectral function at the lowest available temperature \cite{Petreczky:2003js}. This is not needed here and therefore the ratios in Fig.~\ref{fig:G} (below) are a proper reflection of the presence of the temperature-dependent medium.

\begin{figure}[t]
\begin{center}
 \includegraphics*[width=\textwidth]{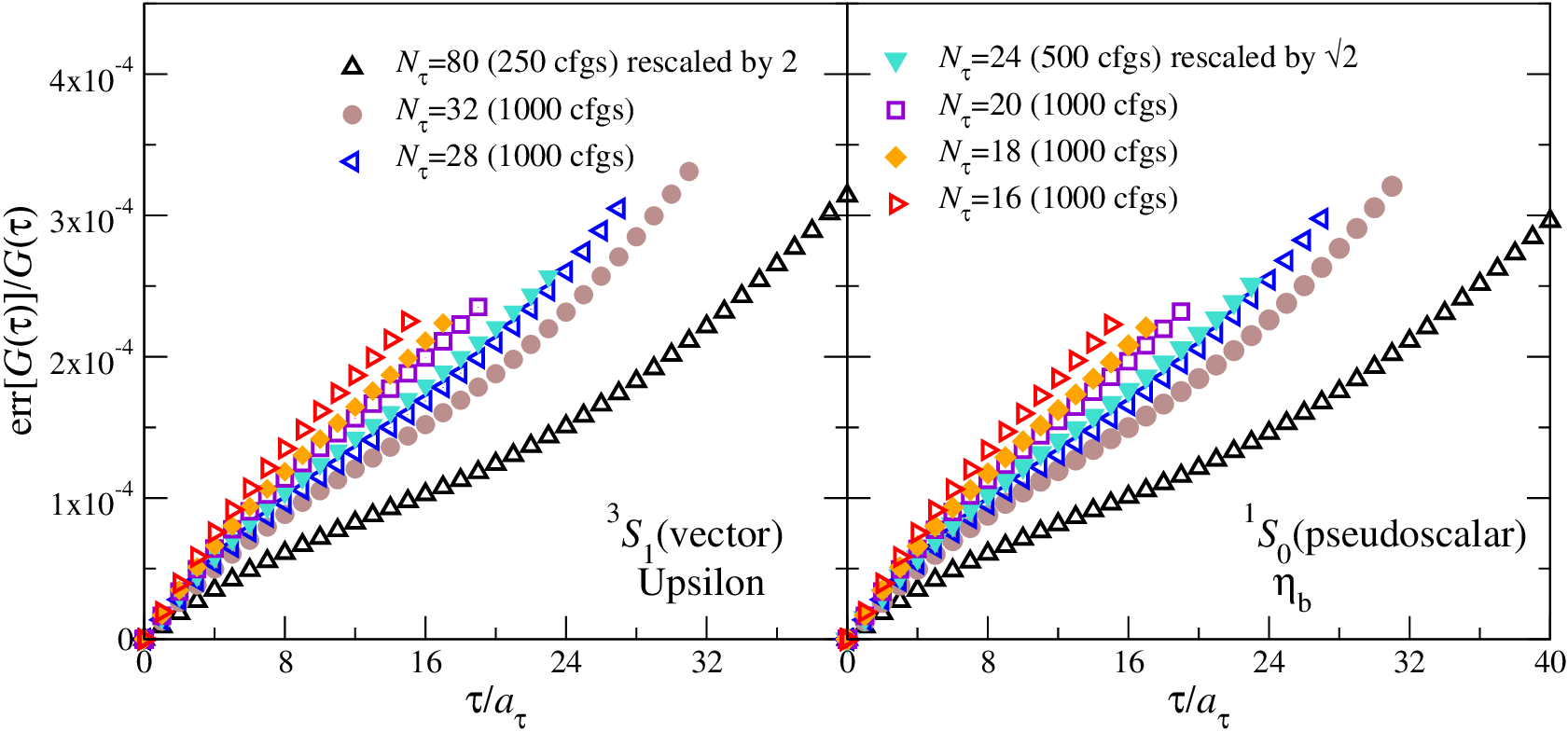}
\end{center} 
 \caption{
Relative error  err[$G(\tau)$]$/G(\tau)$ as a function of euclidean time in the vector (left)  and the pseudoscalar (right) channels. The data at $N_\tau=80$ and $24$ are rescaled to take into account the lower number of configurations. 
   }
 \label{fig:relerror}
 \end{figure}

The statistical errors are small and not visible in the plots discussed above. To illustrate this, we show the statistical relative error, i.e., err[$G(\tau)$]$/G(\tau)$, as a function of euclidean time in Fig.\  \ref{fig:relerror}. At all but two temperatures, there are 1000 configurations available. To take this into account, we rescaled the $N_\tau=80$ data by a factor 2 (250 configurations) and the $N_\tau=24$ data by a factor $\sqrt{2}$ (500 configurations). We then observe that the relative error is of the order of $10^{-4}$ and that it increases as the temperature of the quark-gluon plasma is increased, indicating  larger thermal fluctuations in the hot phase.

\begin{table}[h]
\begin{center}
\begin{tabular}{lccc}
 \hline\hline 
 \multicolumn{1}{l}{state} & \multicolumn{1}{c}{$a_\tau\Delta E$} 
 & \multicolumn{1}{c}{Mass (MeV)} & \multicolumn{1}{c}{Exp.\ (MeV) \cite{Nakamura:2010zz} } \\ 
 \hline
 1$^1S_0 (\eta_b)    $ & 0.118(1) &  9438(8)  &  9390.9(2.8)  \\
 2$^1S_0 (\eta_b(2S))$ & 0.197(2) & 10019(15) &  -            \\
 1$^3S_1 (\Upsilon)  $ & 0.121(1) &  9460$^*$ &  9460.30(26)  \\
 2$^3S_1 (\Upsilon') $ & 0.198(2) & 10026(15) & 10023.26(31)  \\
  1$^1P_1 (h_b)       $ & 0.178(2) &  9879(15) & 9898.3$\pm 1.1^{+1.0}_{-1.1}$~\cite{Adachi:2011ji} \\
  1$^3P_0 (\chi_{b0}) $ & 0.175(4) &  9857(29) &  9859.44(42)(31) \\
  1$^3P_1 (\chi_{b1}) $ & 0.176(3) &  9864(22) &  9892.78(26)(31) \\
  1$^3P_2 (\chi_{b2}) $ & 0.182(3) &  9908(22) &  9912.21(26)(31)  \\
 \hline
 \end{tabular}
 \caption{Zero temperature bottomonium spectroscopy. The 
$1^3S_1 (\Upsilon)$ state is used to set the scale. Here we concentrate on the $\Upsilon$ and $\eta_b$ states.
}
\label{tab:bindE}
\end{center}
\end{table}

From the data at the lowest temperature, we extract the masses of the
ground states and the first excited states using standard exponential
fits. The results are summarised in Table \ref{tab:bindE} and were
already presented in Ref.\ \cite{Aarts:2010ek}.\footnote{In Ref.\
\cite{Aarts:2010ek} a temporal lattice spacing of $a_\tau^{-1}=7.23$
GeV rather than 7.35 GeV was used, correcting this results in a small
change in the mass predictions in the third column.}  In NRQCD, all
energies are determined only up to an additive constant, $E=E_0+\Delta
E$.  Taking the $\Upsilon(1S)$ mass from the Particle Data Book
\cite{Nakamura:2010zz} to set the scale, we find $E_0=8.57$ GeV.

\section{Spectral functions}
\label{sec:spec}

We extract spectral functions from the euclidean correlators presented above using the Maximum Entropy Method (MEM) \cite{Asakawa:2000tr}.  A straightforward inversion of Eq.~(\ref{eq:Gnr}) is not possible, since euclidean correlators are determined numerically at a finite number of points, whereas spectral functions are in principle continuous functions of the energy $\om$. Using the ideas of Bayesian probability theory, one can construct the most probable spectral function by maximizing the conditional probability $P[\rho|DH]$, where $D$ indicates the data and $H$ some additional prior knowledge, encoded in a default model. In this section we present the results, while a discussion of the systematic uncertainties  is given in the next section. We only consider spectral functions at zero spatial momentum.

\begin{figure}[h]
\begin{center}
 \includegraphics*[width=7cm]{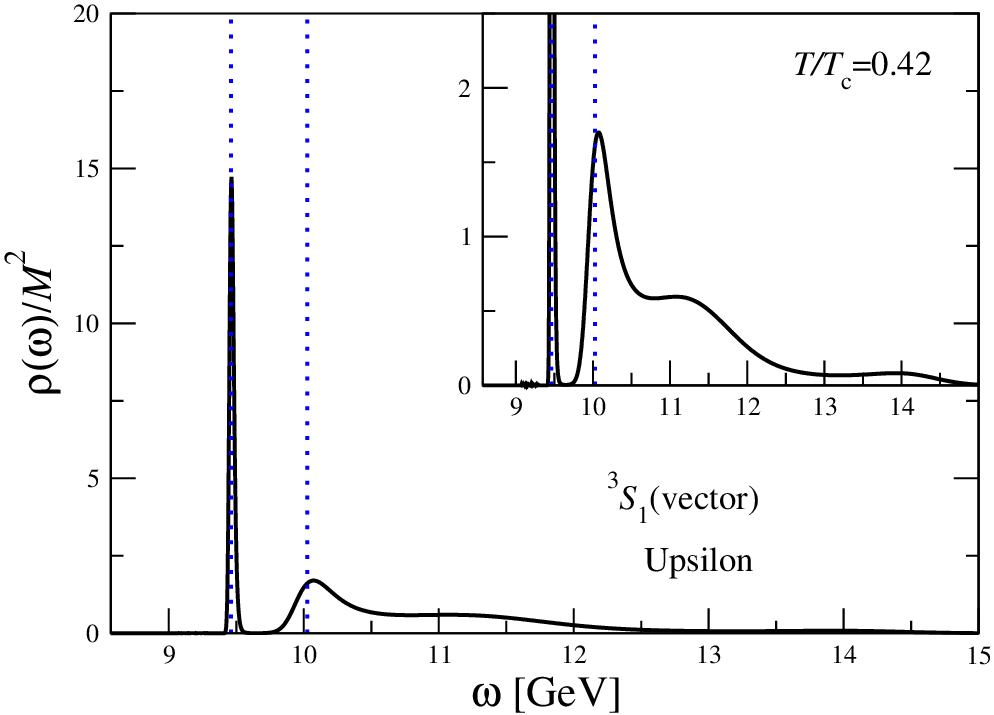}
 \includegraphics*[width=7cm]{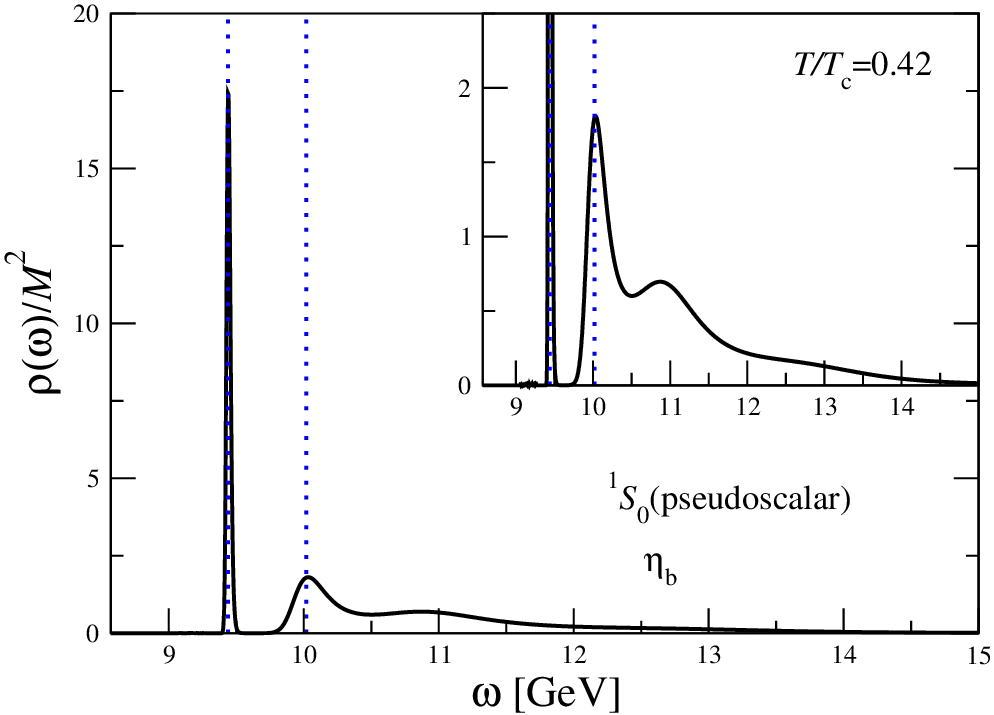}
\end{center} 
 \caption{
Spectral functions $\rho(\omega)$, normalised with the heavy quark mass,  as a function of energy at the lowest temperature
in the vector ($\Upsilon$)  and the pseudoscalar ($\eta_b$) channels. The vertical lines indicate the positions of the ground and first excited state obtained via standard exponential fits. The insets show a close-up. 
}
 \label{fig:rho80}
 \end{figure}

The results for the spectral functions at the lowest temperature
are given in Fig.\ \ref{fig:rho80}. The vertical lines indicate the
position of the ground state and first excited state from Table
\ref{tab:bindE}. We observe that the ground state appears as a very
narrow peak in the spectral function, while the first excited state is
broader and overlaps with structure at larger energy. The third
feature, visible in the inset, is presumably a combination of higher
excited states and lattice artefacts (see Appendix
\ref{sec:free}).\footnote{Note in particular that it cannot be
identified with the second excited state $\Upsilon(3S)$, with a mass
of 10.3552 GeV \cite{Nakamura:2010zz}.  } We note that quadruple
precision is required for the MEM inversion, due to the large number
of points ($N_\tau=80$) and the exponential fall-off.

\begin{figure}[!t]
\begin{center}
  \includegraphics*[width=14cm]{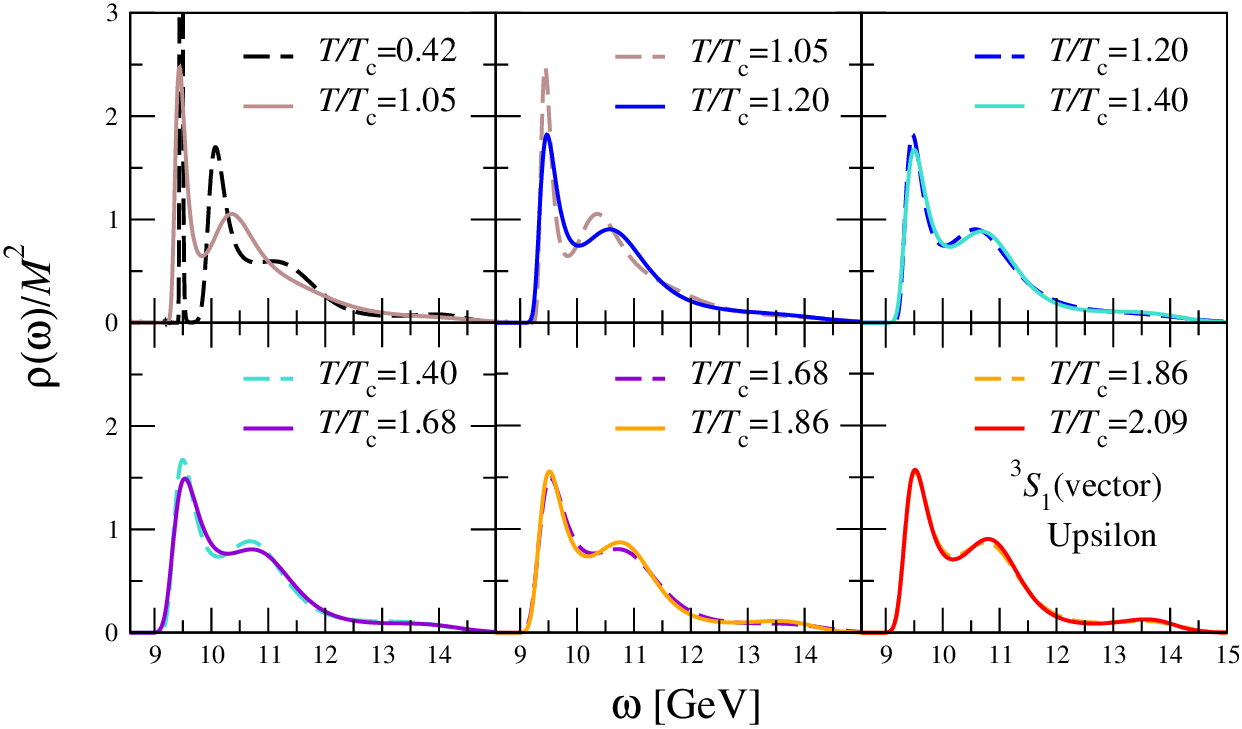} 
  \includegraphics*[width=14cm]{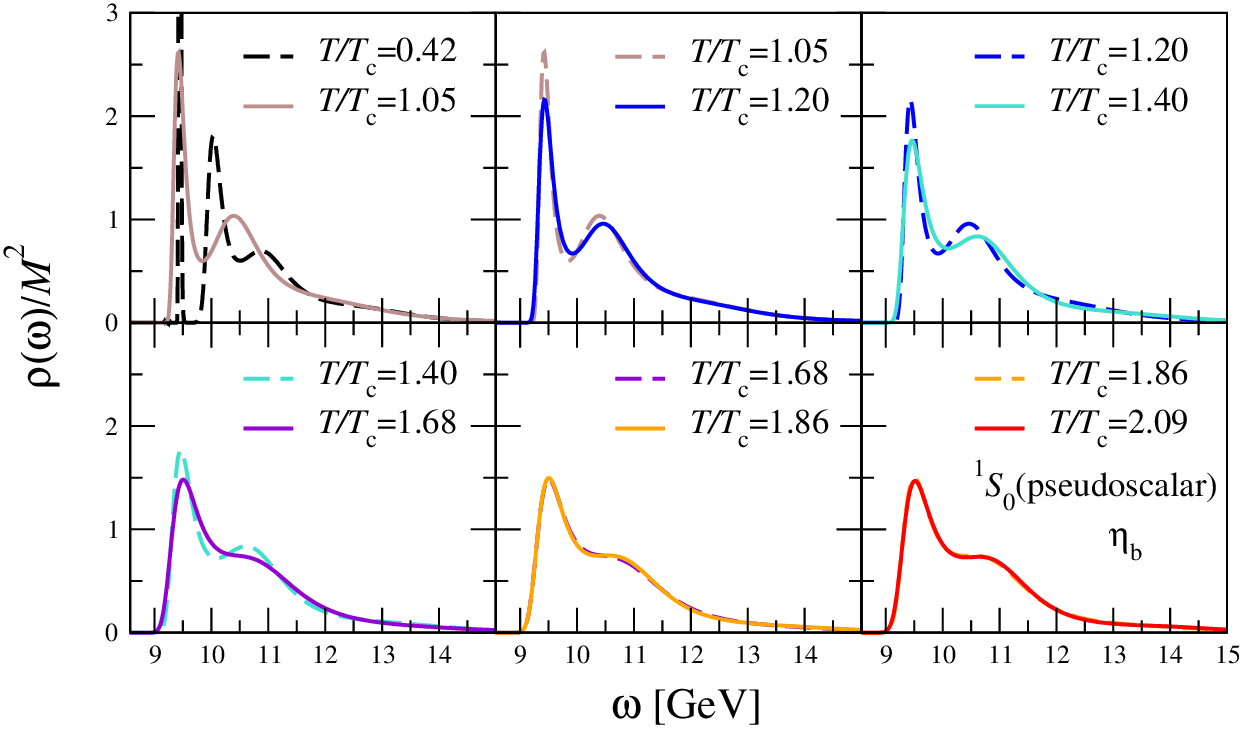} 
 \end{center}
 \caption{ Spectral functions $\rho(\omega)$, normalised with the
heavy quark mass, in the vector ($\Upsilon$) channel  (upper panel) 
and in the pseudoscalar ($\eta_b$) channel (lower panel)
for all temperature available. The subpanels are ordered from cold (top left) to
hot (bottom right). Every subpanel contains two adjacent temperatures to
facilitate the comparison.  }
\label{fig:rho_all_upsilon}
\end{figure}

The main result of this paper is the spectral functions at the various
temperatures, shown in Fig.\ \ref{fig:rho_all_upsilon} for the vector
($\Upsilon$) channel (upper panel) and the pseudoscalar ($\eta_b$) channel
(lower panel).  Every panel contains two adjacent temperatures, from the
coldest ($T/T_c=0.42$) in the top left to the hottest ($T/T_c=2.09$)
in the bottom right.  In each panel, the lower temperature is depicted
with a dashed line and the higher one with a solid line.  As the temperature is
increased, we observe that the ground state peak remains visible, even
though it broadens and reduces in height.  The excited states become
less pronounced as the temperature increases and are no longer
discernible as a separate peak between $1.4 \lesssim T/T_c\lesssim
1.68$.\footnote{We also note that the second peak immediately above
$T_c$ is presumably a combination of the first excited state and other
features.} This can be interpreted as the survival of the $1S$ states,
but a melting or suppression of the excited states. From the analysis
in Sec.\ \ref{sec:sys}, we note here that we consider the results for
the spectral functions to be robust for all temperatures, with the
exception of the two highest temperatures where uncertainties due to
the limited statistics and euclidean time range remain.

We note that the area under the curve is determined by the source\footnote{The point source is defined to be unity for each of the upper two-component spinor indices as well as for each of the colour indices.}
at $\tau=0$ and the spectral relation,
\be
\int\frac{d\om}{2\pi}\,\rho(\om, \pv=\vecnul) = \int d^3x\, G(\tau=0, \xv) = \int d^3x\, S(\xv),
\ee
and is independent of the temperature.

\begin{figure}[!t]
\begin{center}
  \includegraphics*[height=5cm]{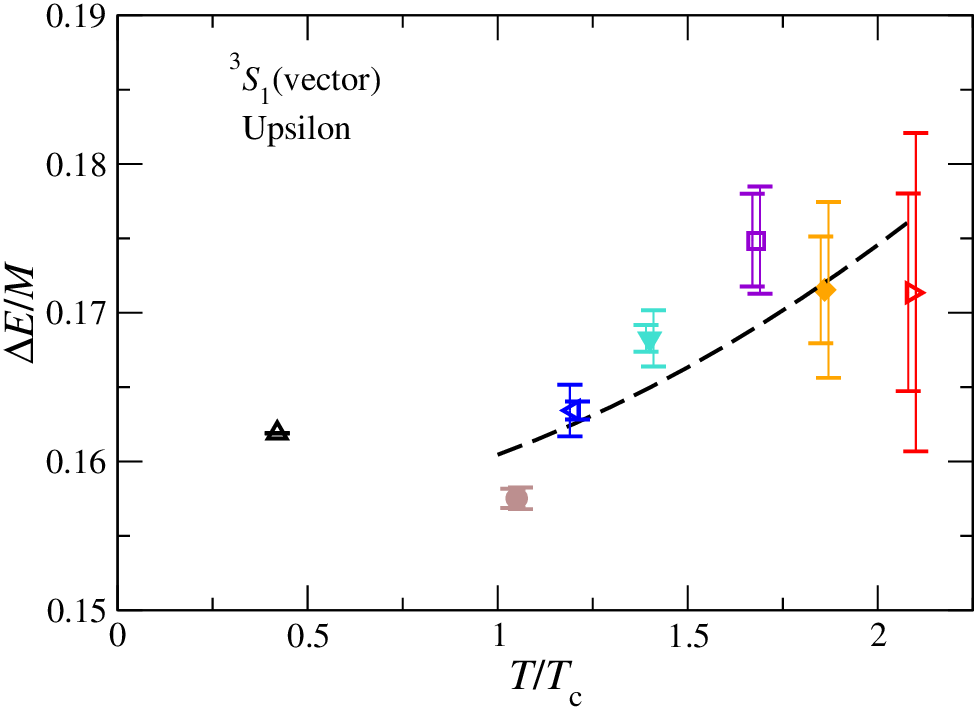}
  \includegraphics*[height=5cm]{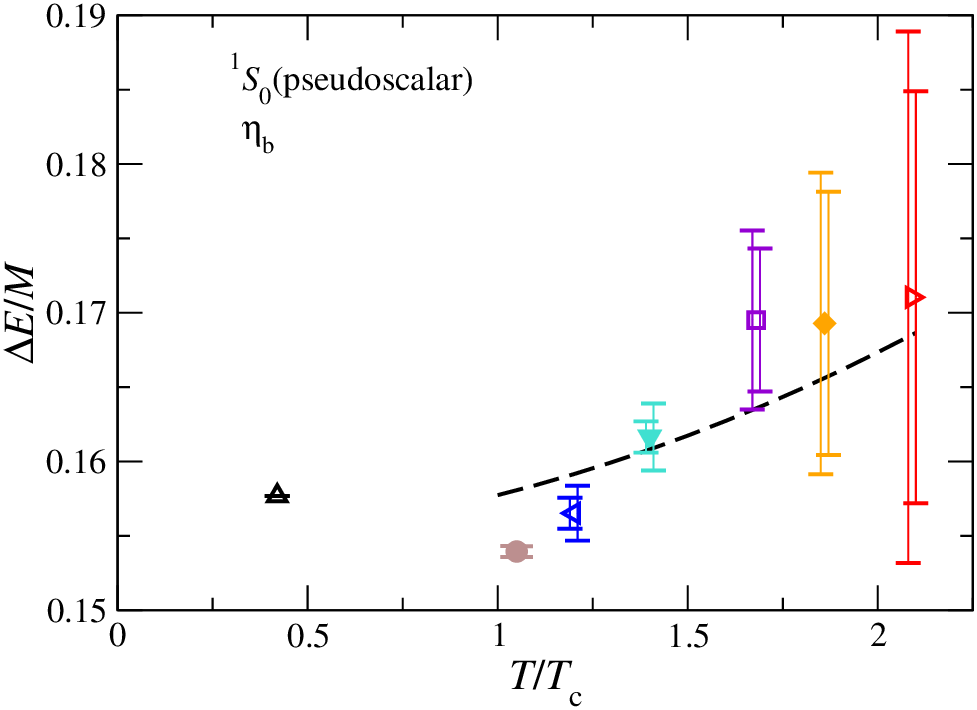}
  \includegraphics*[height=5cm]{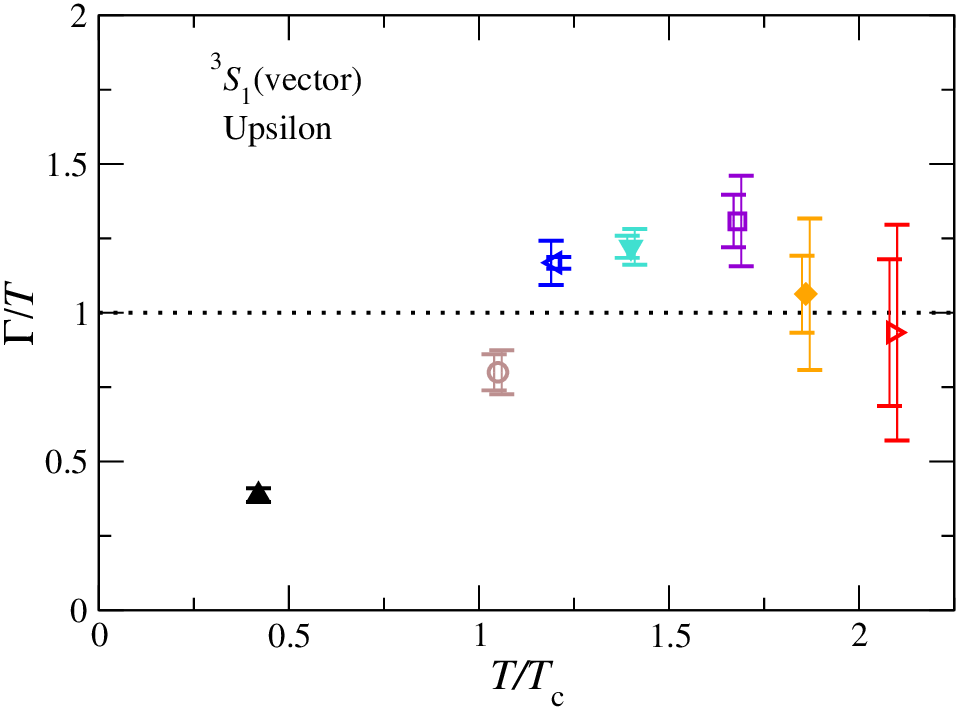}
  \includegraphics*[height=5cm]{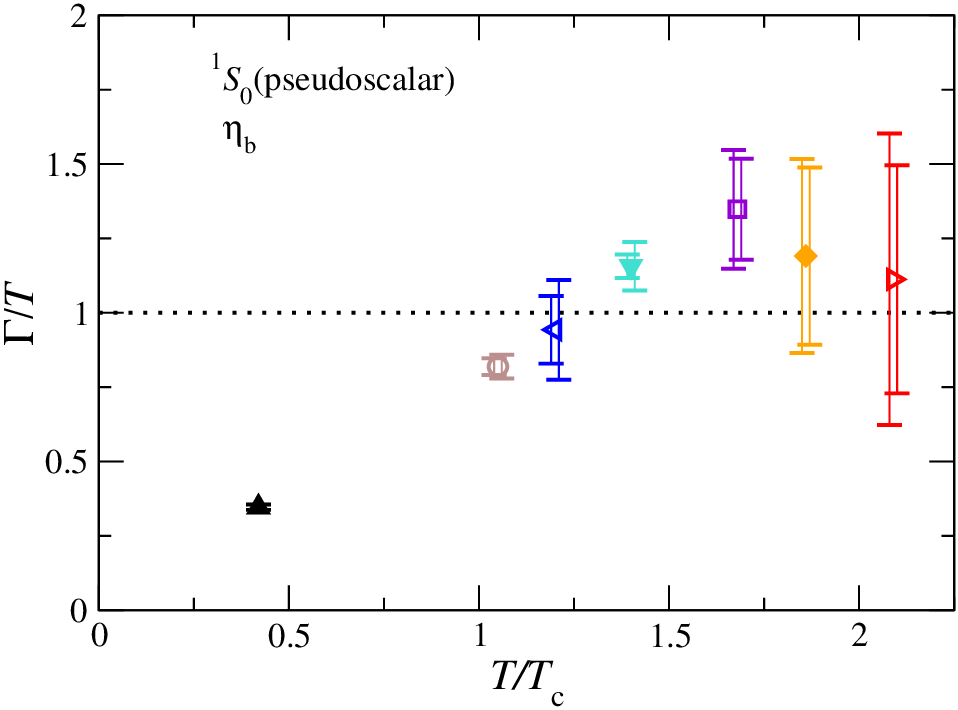}
 \end{center} 
 \caption{
Position of the ground state peak $\Delta E$, normalised with the
heavy quark mass (upper panels), 
and the upper limit on the width of the ground state peak, normalised
with the temperature (lower panels), 
as a function of $T/T_c$ in the vector ($\Upsilon$)  and the
pseudoscalar ($\eta_b$) channels.
The error bars denote the systematic uncertainty with the left error
bars representing the error from the finiteness of the last time in the fitting window,
$\tau_2$, and the right error bar
representing the error from the finite statistics (see Sec.\ \ref{sec:sys}).
The lines in the upper plots indicate expected analytical behaviour assuming weak coupling above $T_c$.
}
 \label{fig:mass}
  \end{figure}

The peak position $E$ and width $\Gamma$ of the ground states can be
extracted by fitting the peaks to a Gaussian function. We fit to the left side of the peak, to avoid contamination
from the features at larger $\om$.  In Fig.\ \ref{fig:mass} (top
panel) we show the temperature dependence of the mass shift $\Delta
E$, normalised with the heavy quark mass. Recall that in NRQCD 
only energy differences can be determined, and that the total energy
is $E=E_0+\Delta E$, where $E_0=8.57$ GeV in our case.  The
temperature dependence of the width is shown in Fig.\ \ref{fig:mass}
(bottom panel).  Note that the width is normalised with the
temperature.  The error bars indicate systematic uncertainties in
extracting the peak position and width from the peaked structure. In
Sec.\ \ref{sec:sys} these uncertainties are discussed in
detail.  Based on this discussion we conclude conservatively  that  the width shown in
Fig.~\ref{fig:mass} is better interpreted as an upper
bound, rather than the width itself.

To see whether these results are reasonable, we now take them at face value and  contrast them with analytic
  predictions derived assuming a weakly coupled plasma. 
According to Ref.\ \cite{Brambilla:2010vq}, the thermal contribution to the width is given, at leading order in the weak coupling and large mass expansion, by
\be
\frac{\Gamma}{T} =  \frac{1156}{81}\as^3 \simeq 14.27\as^3,
  \ee
  i.e., the width increases linearly with the temperature.\footnote{A linearly rising width with temperature was also predicted in Ref.\ \cite{Laine:2006ns}.}
 If we take as an estimate from our results that  $\Gamma/T\sim 1$, we find that
 this corresponds to $\as\sim 0.4$, which is a reasonable result.
In the same spirit  the thermal mass shift is given in Ref.\cite{Brambilla:2010vq} by
\begin{equation} 
 \delta E_{\rm thermal} =  \frac{17\pi}{9}\as\frac{T^2}{M} \simeq
 5.93\as\frac{T^2}{M}. \label{eq:deltaE-pert}
\end{equation}
In these simulations we have $T_c\sim 220$ MeV, $M\sim 5$ GeV.  Taking
these values together with $\as\sim 0.4$ as determined
above, Eq.\ \eqref{eq:deltaE-pert} becomes
\be
 \frac{\delta E_{\rm thermal}}{M} =5.93\as \left(\frac{T_c}{M}\right)^2\left(\frac{T}{T_c}\right)^2 \sim 0.0046\left(\frac{T}{T_c}\right)^2.
 \ee
In order to contrast our results with this analytical prediction, we have compared the temperature dependence of the peak positions to the simple expression
\begin{equation}
 \frac{\Delta E}{M} =c+ 0.0046\left(\frac{T}{T_c}\right)^2, 
\end{equation}
where $c$ is a free paramater. This is shown by the dashed lines in
Fig.\ \ref{fig:mass} (top panel). While we are not in a position to confirm or
rule out the quadratic temperature dependence due to the systematic
uncertainties in the MEM analysis and the fitting of the ground state
peaks, we note that our results are not inconsistent with this. In
particular, the absolute scale of temperature variation seems to be of
the correct order. We also note that the mass just above $T_c$ is reduced with respect to the low temperature one.

\section{Systematics and uncertainties}
\label{sec:sys}

In order  to assess the robustness of the results presented above,  in this section we discuss the main 
systematic uncertainties, namely the dependence on the default model, the choice of $\om_{\rm min}$, the number of configurations and the euclidean time window,  and explain the error estimates of the previous section.
We only show results for the vector channel; the ones in the pseudoscalar channel are similar.

\subsubsection*{Default model}

The MEM procedure includes a ``default model'' $m(\omega)$ which is
used to define the entropy term,
 \be
 \label{eq:S}
 S = \int_{\om_{\rm min}}^{\om_{\rm max}} \frac{d\om}{2\pi} \left[ \rho(\om) - m(\om) -
\rho(\om)\ln\frac{\rho(\om)}{m(\om)} \right],
\ee
in the MEM approach.
It is expected that with poor data the spectral function will resemble the default model, since this minimises the entropy,  but that with precise data  the choice of default model becomes irrelevant. The usual procedure is to choose a default
model which has the same functional form as the (continuum) free
spectral function in the channel under consideration. For the nonrelativistic $S$ wave, this
is $m(\omega) = m_0 \sqrt\omega$, see Eq.\ (\ref{eq:rhoS}). 
An alternative is to use a constant default model, $m(\omega)
= m_0$. The
constants $m_0$ can be set by minimising the $\chi^2$ between the
data and the correlation function defined from the
default model,
\[
G_{\rm def}(\tau) =  \int_{\om_{\rm min}}^{\om_{\rm max}}  \frac{d\omega}{2\pi}\, e^{-\omega  \tau} m(\omega).
\]
To illustrate the absence of default model dependence in our analysis, we show here results for the following six default models:
\begin{itemize}
\item
$m(\om) = m_0  \sqrt\omega$, with $m_0/m_0^*= 0.1, 1,10$, 
\item
$m(\om) = m_0$, with $m_0/m_0^*=0.1, 1,10$,
\end{itemize}
where in both cases $m_0^*$ is determined by minimizing $\chi^2$ as described above.

 \begin{figure}[t]
\begin{center}
 \includegraphics*[width=7cm]{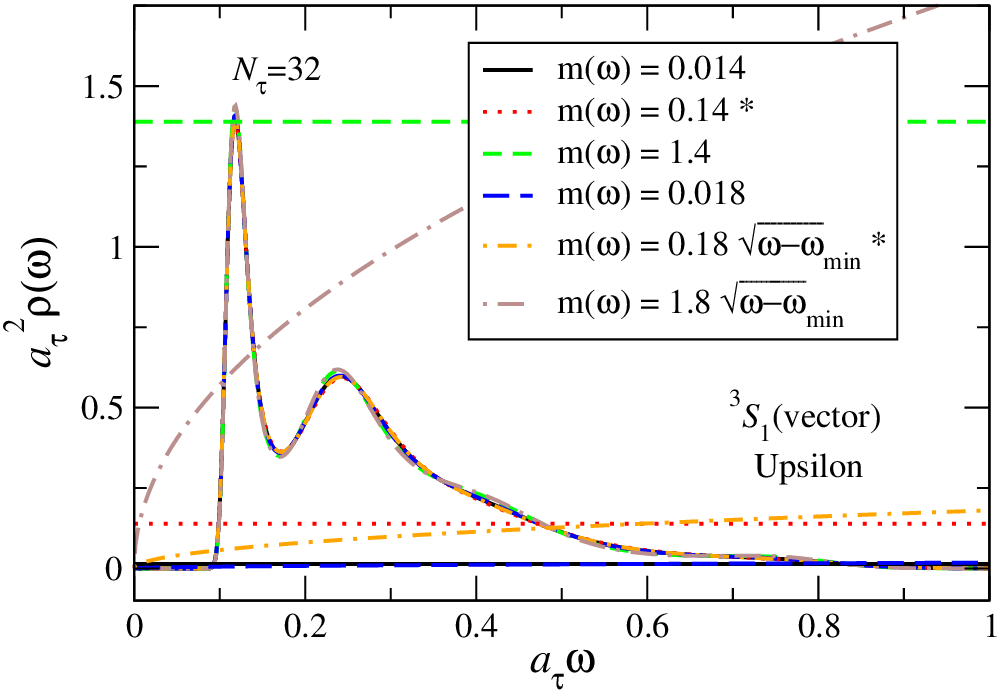}
 \includegraphics*[width=7cm]{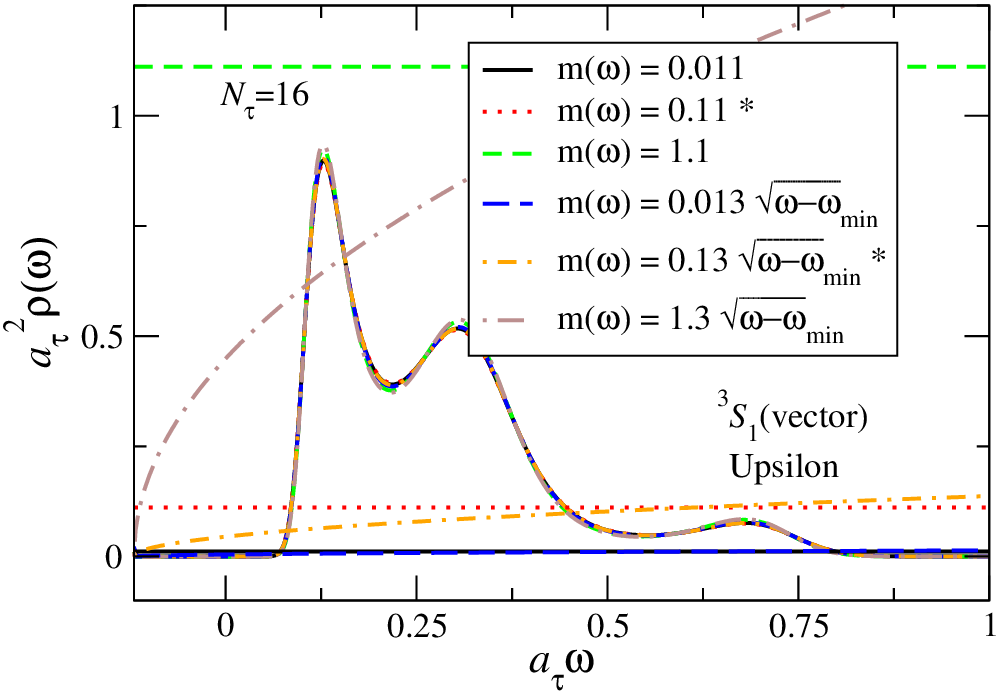}
\end{center} 
 \caption{Dependence of the spectral function on the default model chosen in the MEM analysis 
 for $N_\tau = 32$ (left) and $16$ (right),  in the vector channel. The default models favoured by the $\chi^2$ minimization are denoted with a $*$. }
 \label{fig:upsilonDefModel}
 \end{figure}

In Fig.\ \ref{fig:upsilonDefModel} we show the resulting six spectral functions, as well as the six default models used, 
at $T/T_c=1.05$ and $2.09$.
We observe that  the spectral functions do not resemble the default models and that 
there is almost no variation in the spectral functions, even with default models varying over two orders of magnitude. 
We conclude that there is no significant default model dependence in this analysis, even at the highest temperature.

\subsubsection*{Energy window}

Since in NRQCD the heavy quark mass scale is removed, the additive normalization of the energy scale has to be determined by comparing to a physical state. For this reason, setting the lower boundary $\om_{\rm min}=0$ in the spectral relation
 (\ref{eq:Gnr}) is not a priori justified and it might be necessary to allow for the possibility of
negative energies ($\om_{\rm min}<0$)  in the energy integral.
 Indeed, in the high temperature cases we noticed that when $\omega_{\rm min}\sim 0$, 
the MEM spectral functions have a spike in the lowest-energy bin, which we interpret as an unphysical effect. 
Reducing $\omega_{\rm min}$ to negative values reduces this spike until it is essentially
absent when $a_\tau\omega_{\rm min} \sim -0.10$. 
 At the lower temperatures ($N_\tau = 80, 32$),  no spikes were observed with $\om_{\rm min}=0$. 
In Table \ref{tab:MEM} we list the parameters used in the MEM analysis. $N_\om$ denotes the number of points in which the energy interval $\om_{\rm max}-\om_{\rm min}$ is divided.

\begin{table}[t]
\begin{center}
\begin{tabular}{ccrcrc}
\hline\hline 
 \multicolumn{2}{c}{} & 
 \multicolumn{2}{c}{$\Upsilon$ (vector channel)} & 
 \multicolumn{2}{c}{$\eta_b$ (pseudoscalar channel)} 
\\
 \multicolumn{1}{c}{$N_\tau$} &  \multicolumn{1}{c}{$N_\om$} & 
 \multicolumn{1}{c}{$a_\tau\om_{\rm min}$} & \multicolumn{1}{c}{$a_\tau\om_{\rm max}$} &
 \multicolumn{1}{c}{$a_\tau\om_{\rm min}$} & \multicolumn{1}{c}{$a_\tau\om_{\rm max}$} 
\\ 
\hline 
80 & 4000 & $0.00$ 	 &  2.00 &  $0.00$ 	&  2.00 \\
32 & 1000 & $0.00$ 	 &  1.50 & $0.00$ 	&  2.00 \\
28 & 1000 & $-0.04$ &  1.46 & $-0.08$ 	&  1.92 \\
24 & 1000 & $-0.10$ &  1.40 & $-0.10$ 	&  1.90 \\
20 & 1000 & $-0.10$ &  1.40 & $-0.10$ 	&  1.90 \\
18 & 1000 & $-0.10$	 &  1.40 & $-0.10$ 	&  1.90 \\
16 & 1000 & $-0.12$ &  0.88 & $-0.12$ 	&  1.88 \\
\hline
\end{tabular}
 \caption{Details of the parameters used in the MEM analysis. Note that $\Delta\om=\left(\om_{\rm max}-\om_{\rm min}\right)/N_\om$, and that $a_\tau\om=0$ and $2$  correspond to 8.57 GeV and 23.3 GeV respectively.
 }  
\label{tab:MEM}
\end{center}
\end{table}


\subsubsection*{Number of configurations}

\begin{figure}[t]
\begin{center}
 \includegraphics*[width=7cm]{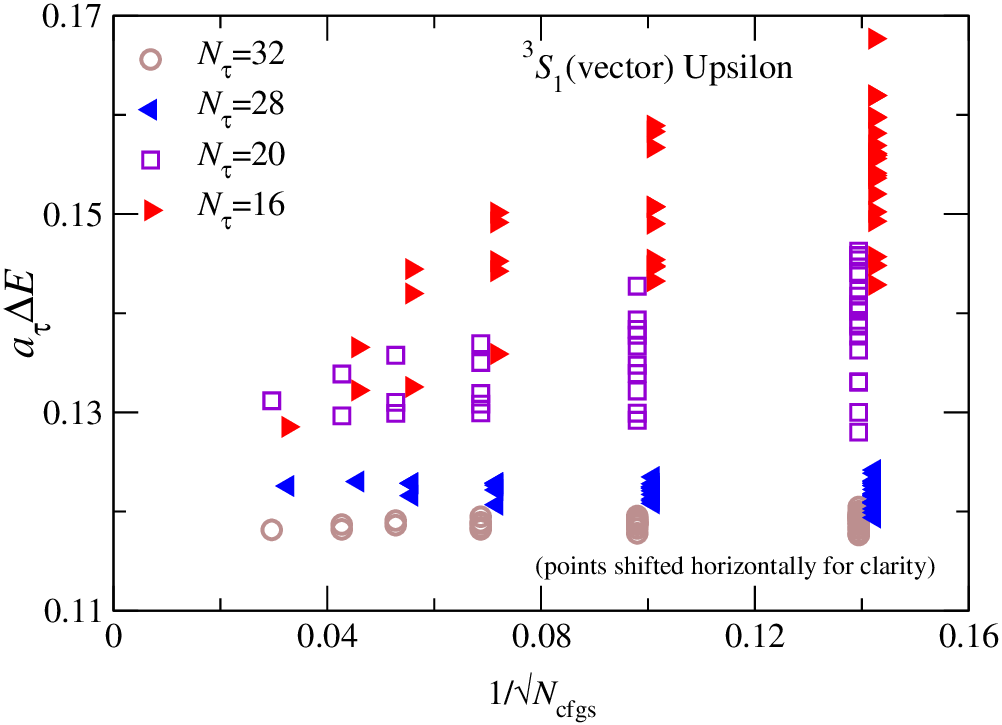}
 \includegraphics*[width=7cm]{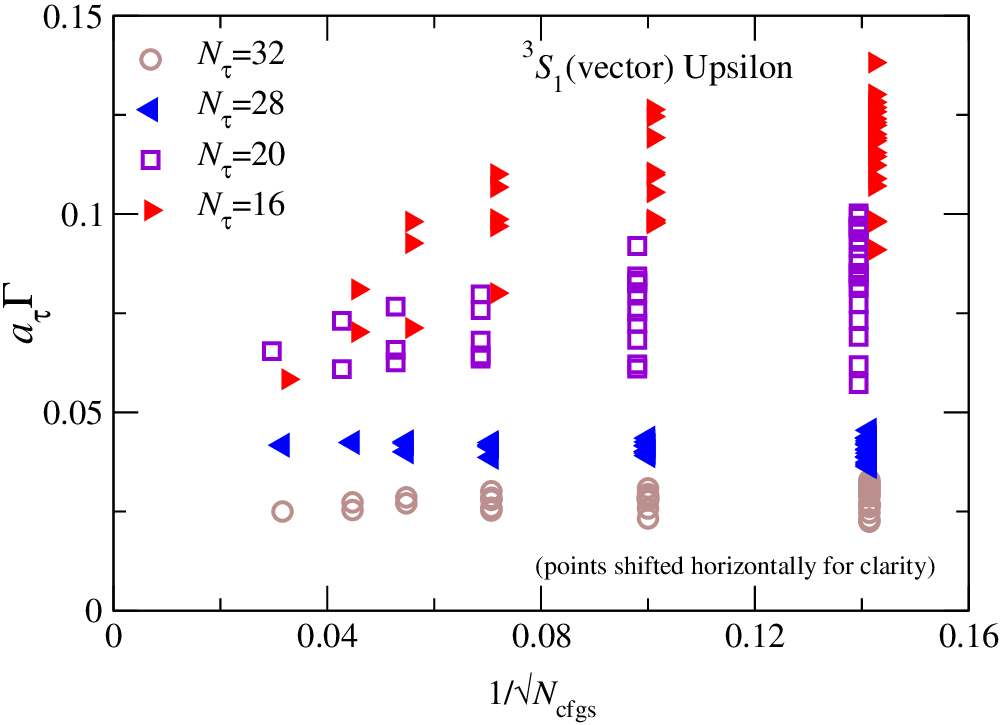}
\end{center} 
 \caption{Dependence of the position ($\Delta E$) and width ($\Gamma$) of the
   ground state peak extracted from the spectral function on the number
   $N_{\rm cfgs}$ of configurations used, in the vector channel.  }
 \label{fig:upsilonNcfgs}
 \end{figure}

In Fig.\ \ref{fig:upsilonNcfgs} we illustrate how the position and
width of the ground state peak depend on the number of configurations
used in the MEM analysis. For clarity, we only show results for a
selection of $N_\tau$ values.  At each temperature, we divide the total
number of configurations (1000) into 20 groups of 50, 10 groups of 100,
5 groups of 200, 3 groups of 333, and 2 groups of 500, and compute the
spectral function for each of them. The resulting peak position and
width are shown as a function of $1/\sqrt{N_{\rm cfgs}}$. At the lower
temperatures ($N_\tau=32,28,24$) we observe that both are essentially
independent of the number of configurations used and the results are
stable. At the higher temperature ($N_\tau=20$) there is a larger
spread for the low-statistics results, but the average results are
again quite stable, with at most only a slight decrease in the
width and mass as the statistics are increased. It is only at the
highest temperatures 
($N_\tau=18, 16$) that there is a clear decrease in the position and width
with increased statistics.  We conclude that the statistical error
due to the finiteness of the ensemble does not prevent us from
determining the ground state mass and width, at all but the highest
two temperatures.

To translate the dependence on the number of configurations into an uncertainty in the position and width of the
ground state, we proceed as follows. The central value is taken from the analysis
with 1000 configurations.  The statistical error due to the
finite number of configurations is defined by taking the
difference between the central value and the value obtained with 500
configurations which is furthest from the central value. This
statistical error is shown as the right-hand error bar in
Fig.\ \ref{fig:mass}.

\subsubsection*{Euclidean time window}

\begin{figure}[t]
\begin{center}
 \includegraphics*[width=7cm]{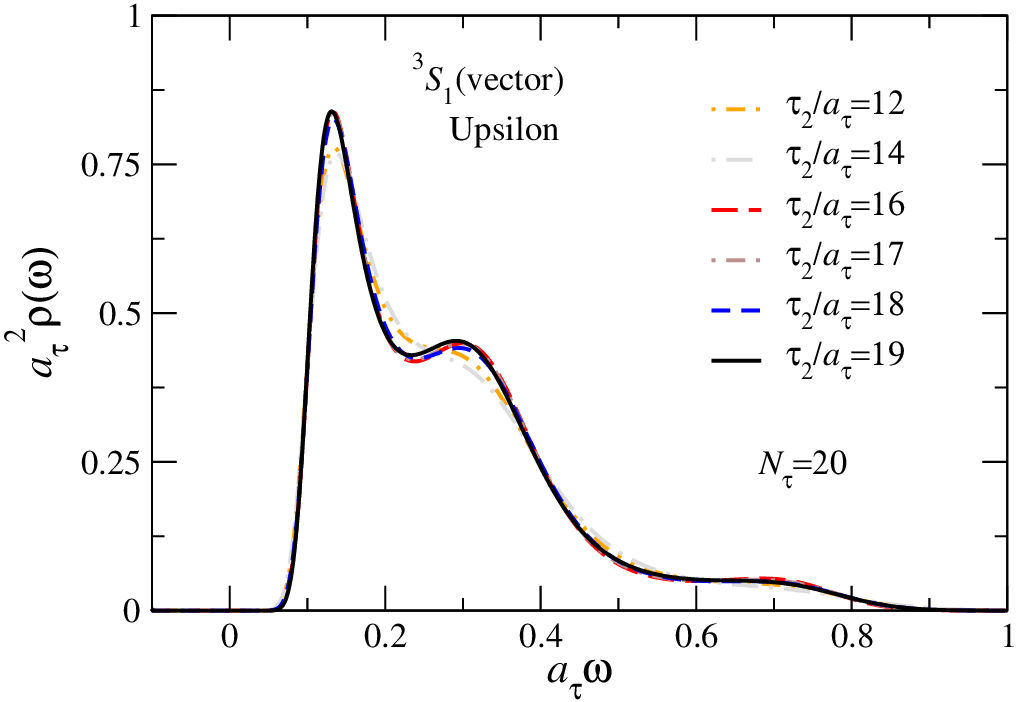}
 \includegraphics*[width=7cm]{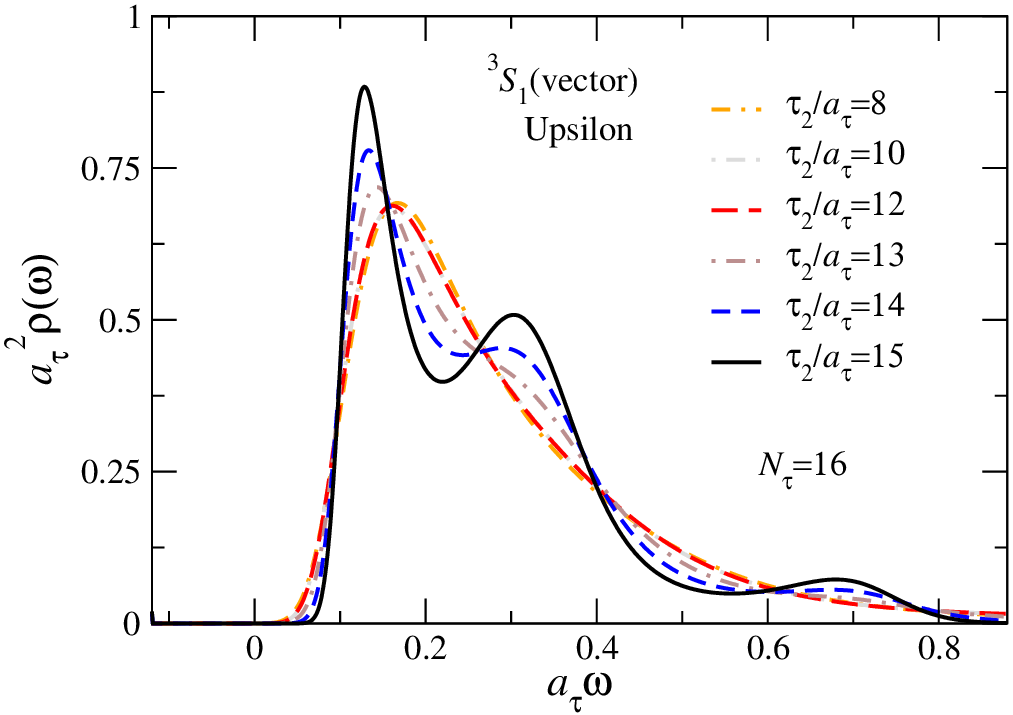}
\end{center} 
 \caption{Dependence of spectral functions on the euclidean time window $a_\tau=\tau_1\leq\tau\leq\tau_2$, for several values of $\tau_2$ for $N_\tau=20$ (left) and 16 (right), in the vector channel. 
   }
 \label{fig:upsilont2rho}
\begin{center}
 \includegraphics*[width=7cm]{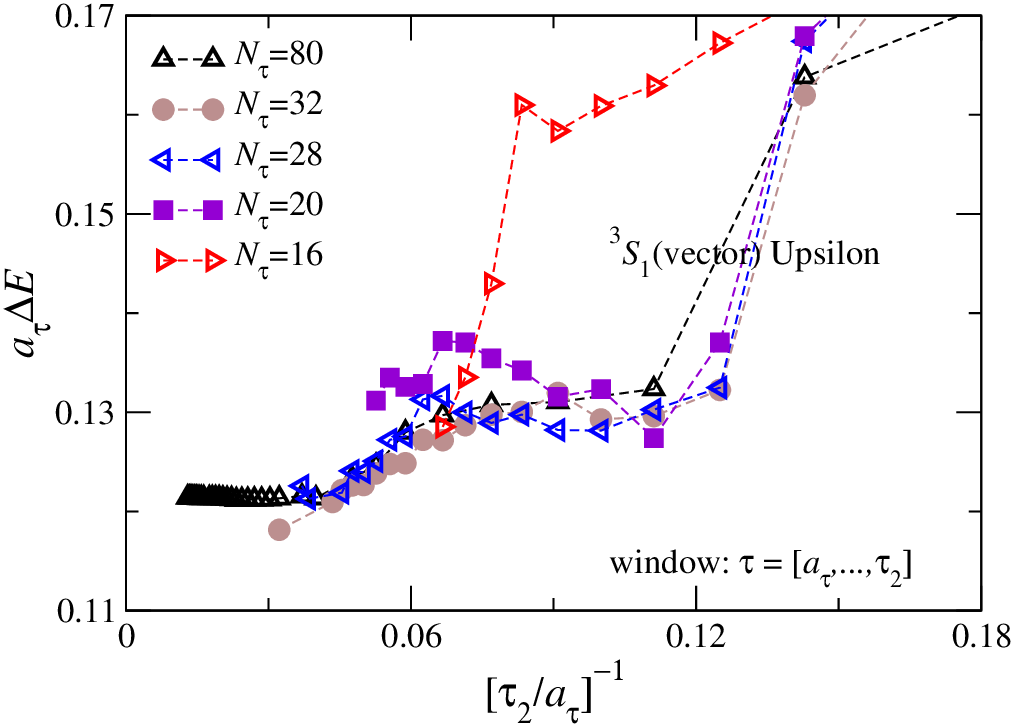}
 \includegraphics*[width=7cm]{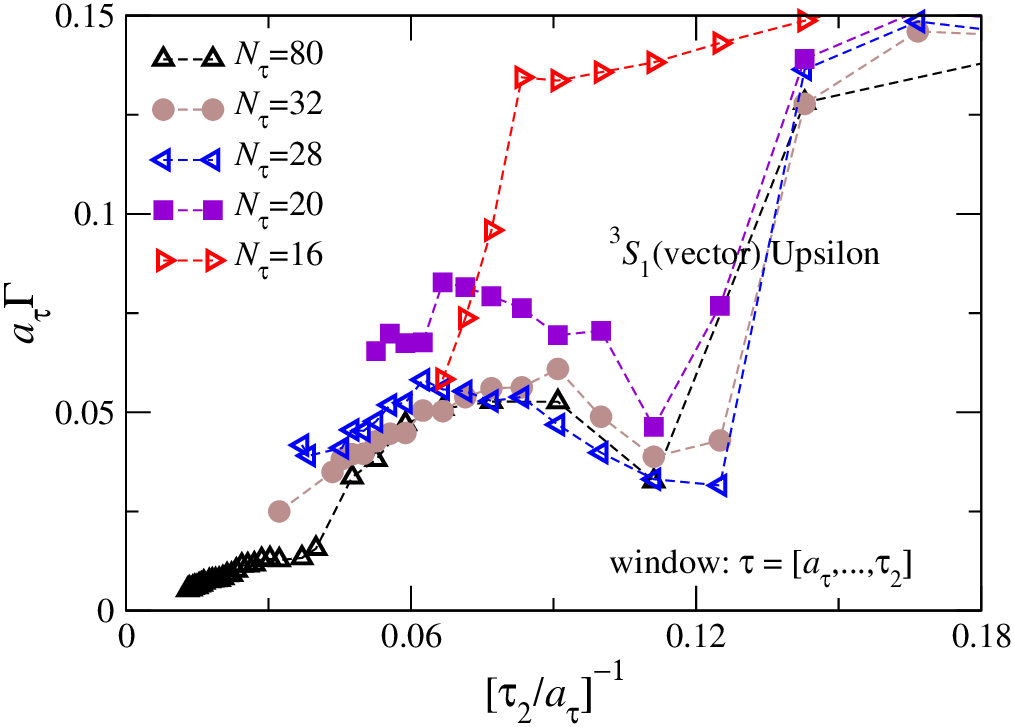}
\end{center} 
 \caption{Dependence of the position ($\Delta E$, left) and width ($\Gamma$, right) of the ground state peak extracted from the spectral function on the euclidean time window $a_\tau=\tau_1\leq \tau\leq \tau_2$ used in the MEM analysis, in the vector channel. 
   }
 \label{fig:upsilont2}
 \end{figure}

The euclidean time window included in the MEM analysis is $\tau_1\leq
\tau \leq \tau_2$, with the initial time equal to the first time
slice, i.e.\ $\tau_1=a_\tau$.\footnote{Varying $\tau_1$ within reason
did not have a significant effect.}  In Fig.\ \ref{fig:upsilont2rho}
we show the dependence of the constructed spectral functions on
$\tau_2$ for $N_\tau=20$ (left) and $N_\tau=16$ (right), varying
$\tau_2$ from a small value up to $N_\tau-1$. Both cases are at high
temperature, $T/T_c=1.68$ and $2.09$ respectively. We observe that at
$N_\tau=20$ the result is stable and the limitation of having only a
finite number of points in the temporal direction does not appear to
be a problem. At the highest temperature, on the other hand, we note
that the results have not yet stabilised, so that the uncertainty in
constructing the spectral function is larger.

In order to quantify this, we have determined the $\tau_2$ dependence
of the position and peak of the ground state at all temperatures. The
result is shown in Fig.\ \ref{fig:upsilont2} for a selection of
$N_\tau$ values.  For the lowest temperature ($N_\tau=80$), we observe
that $\Delta E$ is stable and consistent with the result from the
standard fits directly to the euclidean correlator. Moreover, there is
evidence that the width $\Gamma$ decreases towards zero as $\tau_2$
increases.  The position and width at $N_\tau=32, 28$, or $T/T_c=1.05, 1.20$,
appear to behave as in the hadronic phase, except for a possible
flattening out of the width for $N_\tau=28$ at large $\tau_2$.  A
similar behaviour is found for $N_\tau=24$, with a clearer tendency
for both the position and width to flatten out at large $\tau_2$.  On the
other hand, at $N_\tau=20$ both
position and width are consistently above their low-temperature
values, and may reach a plateau for large
$\tau_2$, consistent with Fig.\ \ref{fig:upsilont2rho}.  For
$N_\tau=18$ and 16, no stability is seen.
For $N_\tau=16$, it follows from Fig.\ \ref{fig:upsilont2rho} that
when $\tau_2$ is too small, the ground state and features at larger
energies overlap and therefore the ground state peak cannot be
resolved. At $\tau_2/a_\tau\gtrsim 13$, two structures become
visible, which explains the rapid drop of the position and peak.  We
conclude therefore that at most temperatures above $T_c$ the position
and width of the groundstate peak appear to be stable, except at the
highest two temperatures where substantial dependence on $\tau_2$ remains.

To translate the finiteness of $\tau_2$  into a systematic uncertainty on the position and width of the ground state, we adopt the following procedure. The central value
is taken from the analysis with $\tau_2 = N_\tau-1$.  The systematic
error due to a finite $\tau_2$ is defined by taking the difference
between the central value and the value obtained with $\tau_2 = N_\tau
-2$. This systematic error is shown as the left-hand error bar in
Fig.\ \ref{fig:mass}.

\subsubsection*{Bryan's approach}

\begin{figure}[t]
\begin{center}
 \includegraphics*[width=7cm]{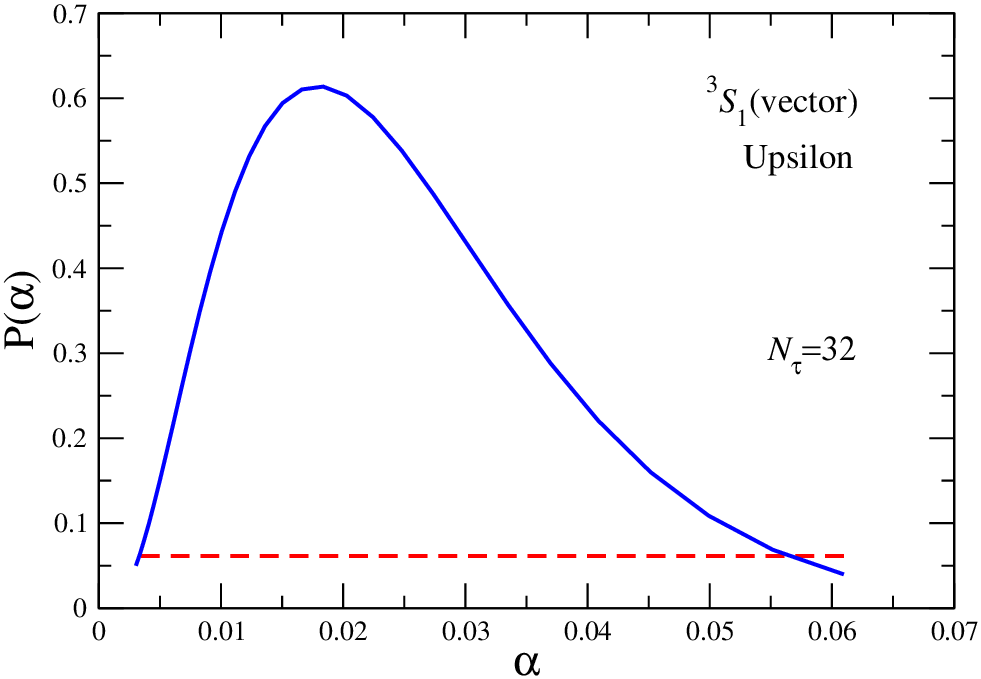}
 \includegraphics*[width=7cm]{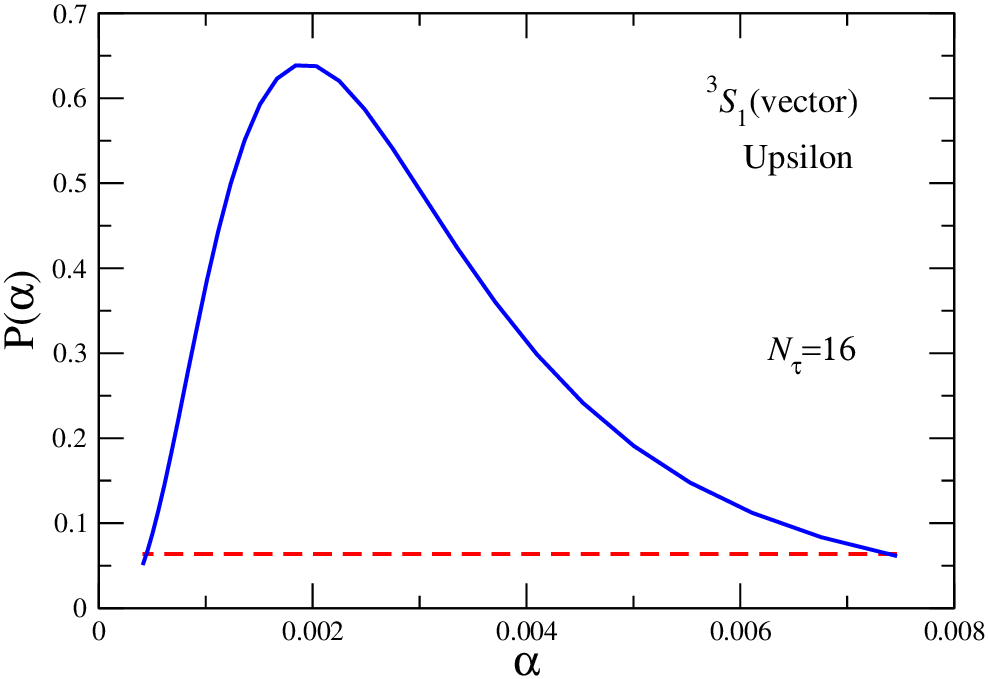}
\end{center}
 \caption{
The probability $P(\alpha)$ used in Bryan's approach for
$N_\tau = 32$ (left) and 16 (right). The horizontal lines
indicate the interval used in the integral in Eq.~(\ref{eq:prob}).
}
\label{fig:prob}
\end{figure}

In the MEM approach, Bayes' theorem implies that the entropy term $S$ is
balanced against the usual maximum likelihood $\chi^2$. The
function $Q$ to be maximized is 
\be
Q = \alpha S - \chi^2,
\ee
where $S$ is defined in Eq.\ (\ref{eq:S}) and $\alpha$ is a normalisation constant which is to be
determined. In Bryan's approach \cite{Bryan} the spectral
function $\rho_\alpha(\om)$ is calculated for each value of $\alpha$ and the
final spectral function is obtained by performing the convolution integral,
\begin{equation}
\rho_\text{final}(\omega) = \int d\alpha\, P(\alpha) \rho_\alpha(\omega).
\label{eq:prob}
\end{equation}
Here $P(\alpha)$ is the probability that $\alpha$ is chosen correctly.

In Fig.\ \ref{fig:prob}, the probability $P(\alpha)$ is plotted for
the $\Upsilon$ channel for $N_\tau=32$ and 16. In practice, the
limits in the integral in Eq.\ (\ref{eq:prob}) are taken such that
$P(\alpha)$ is greater than some fraction of its maximum value $P_{\rm max}$.  In
this case, we use the interval defined as $P(\alpha)/P_{\rm max} > 0.1$ and the resulting
interval is shown in Fig.~\ref{fig:prob} by the horizontal lines. In all cases we find that the probability is
well defined with a clear maximum.

\section{Conclusion}
\label{sec:conclusion}

In this paper we analysed  lattice QCD results for $S$ wave bottomonium correlators (in the $\Upsilon$ and $\eta_b$ channels).  The heavy quarks are treated with NRQCD and propagate through a two-flavour quark-gluon medium at seven temperatures between $0.4T_c$ and $2.1T_c$. Highly anisotropic lattices, with $a_s/a_\tau=6$, are used to maximise the number of time slices available for the analysis.
Spectral functions are constructed with the help of the maximum entropy method.

The use of NRQCD has a number of advantages compared to relativistic
quark dynamics. Since the presence of a nonzero temperature is not
imposed as a thermal boundary condition, twice as many euclidean time
points are available for the analysis compared to the relativistic
case (at the same temperature and lattice spacing). More importantly,
the spectral relation simplifies considerably, removing the problem
with the so-called constant contribution in the
correlator. Physically, it implies that all temperature dependence is
due to the presence of the light-quark--gluon system.

Our main results are spectral functions for the $S$ waves, in the
vector ($\Upsilon$) and pseudoscalar ($\eta_b$) channels. Our results
suggest that the ground state survives up to the highest temperature
we consider, whereas the excited states are suppressed and no longer
visible at temperatures above $1.4T_c$. We have extracted the position
and width of the ground state peaks and found them to be consistent
with analytical results obtained within the EFT framework, at leading
order in the large mass expansion.

Systematic uncertainties have been studied in some detail. For all
temperatures, we found no dependence on the default model used in the
MEM analysis. The position and width of the ground state peaks were
shown to be stable as the number of configurations is increased, for
all except the highest temperatures. Perhaps the biggest uncertainty
comes from the finite number of euclidean time points, but our
analysis suggests that the peak position and width can still be
reliably determined for temperatures up to $1.7T_c$.

We hope that our results will be useful for further EFT and potential
model studies. After constructing a temperature-dependent potential,
one usually computes spectral functions and the corresponding
euclidean correlators. It would be interesting to compare the outcome
of such an exercise with our nonperturbatively determined
correlators. This would in particular be applicable to ratios of
finite-temperature correlators with the zero-temperature one, as in
Fig.\ \ref{fig:G}, to cancel normalization factors and focus on the
temperature dependence.

Finally, we hope that our results will contribute to a further
understanding of the recent experimental results for bottomonium in
heavy ion collisions at the LHC and RHIC.

\section*{Acknowledgments}

We thank Mikko Laine for discussion and clarification.  CA, GA and MPL
thank Trinity College Dublin and the National University of Ireland
Maynooth for hospitality.  We acknowledge the support and
infrastructure provided by the Trinity Centre for High Performance
Computing and the IITAC project funded by the HEA under the Program
for Research in Third Level Institutes (PRTLI) co-funded by the Irish
Government and the European Union.  The work of CA and GA is carried
as part of the UKQCD collaboration and the DiRAC Facility jointly
funded by STFC, the Large Facilities Capital Fund of BIS and Swansea
University.  GA and CA are supported by STFC.  SK is grateful to STFC
for a Visiting Researcher Grant and supported by the National Research
Foundation of Korea grant funded by the Korea government (MEST) No.\
2011-0026688.  SR is supported by the Research Executive Agency (REA)
of the European Union under Grant Agreement number PITN-GA-2009-238353
(ITN STRONGnet) and the Science Foundation Ireland, grant no.\
11/RFP.1/PHY/3201.  DKS is supported in part by US Department of
Energy contract DE-AC02-06CH11357.  JIS is supported by Science
Foundation Ireland grant 08-RFP-PHY1462.

\appendix

\section{Noninteracting lattice spectral functions}
\label{sec:free}

In order to understand the effect of lattice artefacts, it is useful to construct lattice spectral functions in the absence of interactions,  adapting the approach of Refs.\ \cite{Karsch:2003wy,Aarts:2005hg} to lattice NRQCD.

Let us start with free quarks in continuum NRQCD with energy $E_\pv=\pv^2/2M$. The 
correlators for the $S$ and $P$ waves at zero spatial momentum are then of the form 
\cite{Burnier:2007qm}
 \begin{align}
\label{eq:GS}
 G_{S}(\tau)  = &\, 2N_c \int \frac{d^3p}{(2\pi)^3}\, e^{-2E_\pv\tau} 
 = \frac{N_c}{4\pi^{3/2}}\left(\frac{M}{\tau}\right)^{3/2}, \\ 
\label{eq:GP}
 G_{P}(\tau)  = &\, 2N_c \int \frac{d^3p}{(2\pi)^3}\, \pv^2 e^{-2E_\pv\tau}
 = \frac{3N_c}{8\pi^{3/2}} \left(\frac{M}{\tau}\right)^{5/2}, 
\end{align}
 where in the explicit evaluation we did not include an ultraviolet  cutoff, since the integrals are finite for nonzero $\tau$.

This is easily expressed in terms of spectral densities, using
\be
G(\tau) = \int_{\om_{\rm min}}^{\om_{\rm max}} \frac{d\om}{2\pi}\, e^{-\om\tau}\rho(\om),
\ee
yielding
\begin{align}
\label{eq:rhoS}
 \rho_{S}(\om)  = &\, 4\pi N_c \int \frac{d^3p}{(2\pi)^3}\, \delta\left(\om-2E_\pv\right)
 = \frac{N_c}{\pi} M^{3/2} \om^{1/2}\Theta(\om),
  \\ 
\label{eq:rhoP}
 \rho_{P}(\om)  = &\, 4\pi N_c \int \frac{d^3p}{(2\pi)^3}\, \pv^2 \delta\left(\om-2E_\pv \right)
  = \frac{N_c}{\pi} M^{5/2} \om^{3/2}\Theta(\om).
\end{align}
Note that  the minimal energy $\om_{\rm min}=0$ corresponds to twice the heavy quark mass, due to the nonrelativistic approximation, $\sqrt{\pv^2+M^2}\approx M + E_\pv$.
In the presence of a momentum cutoff $|\pv|<\Lambda$, the maximum energy is finite and given by $\om_{\rm max} =\Lambda^2/M$.
Note also that  there is no temperature dependence in the absence of interactions. All temperature effects enter via the propagation through the quark-gluon  system.

\begin{figure}[t]
\begin{center}
 \includegraphics*[width=7cm]{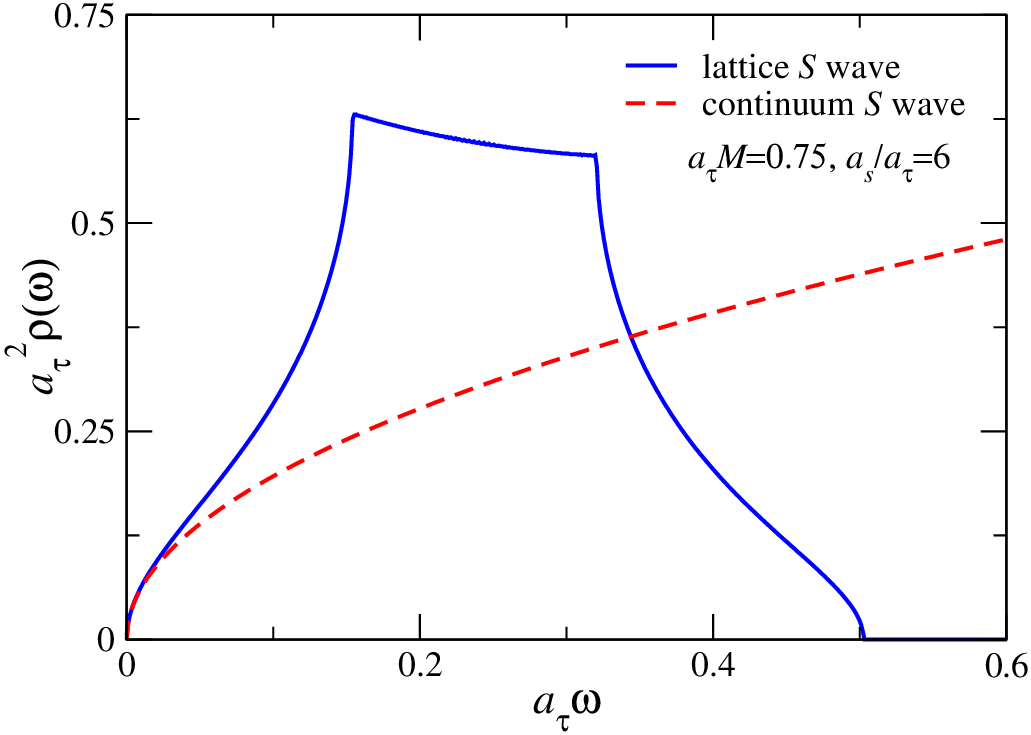}
 \includegraphics*[width=7cm]{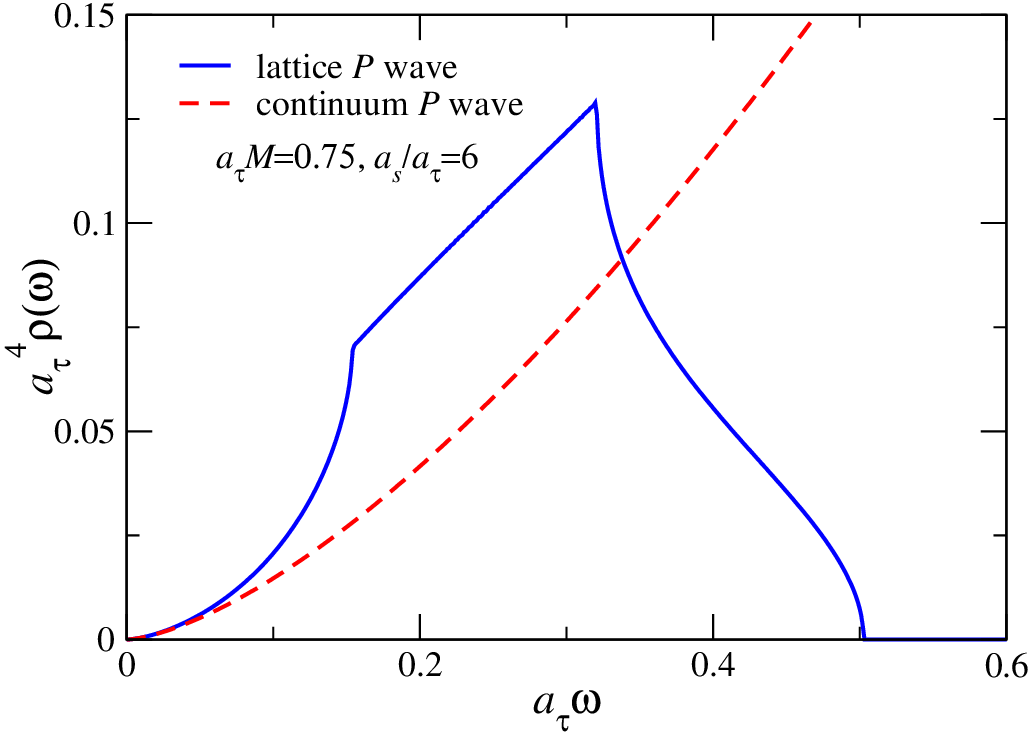}
\end{center} 
 \caption{Lattice spectral functions, in units of the temporal lattice spacing $a_\tau$,  as a function of $a_\tau\om$, for $S$ waves (left) and $P$ waves (right) in anisotropic lattice NRQCD at lowest order, ignoring interactions, using $\xi\equiv a_s/a_\tau=6$ and $a_\tau M=0.75$.
 The dashed lines indicate the continuum spectral functions  in the absence of a cutoff.
     }
 \label{fig:free}
\end{figure}

These results can easily be adapted to the lattice \cite{Karsch:2003wy,Aarts:2005hg}, 
taking into account the lattice dispersion relation and the finite momentum integration over the first Brillouin zone. At lowest order in (unimproved)  NRQCD, the dispersion relation is 
\be
a_\tau E_\pv = -\log\left(1-\frac{\hat\pv^2}{2\xi\hat M}\right),
\ee
where $\hat M=a_s M$, $\xi=a_s/a_\tau$, and 
\be
\hat\pv^2 = 4\sum_{i=1}^3\sin^2\left(\frac{p_i}{2}\right), \;\;\;\;\;\;\;\;\;\; p_i= \frac{2\pi n_i}{N_s},
\;\;\;\;\;\;\;\;n_i=-\frac{N_s}{2}+1, \dots, \frac{N_s}{2},
\ee
with $N_s$ the number of sites in a spatial direction.
The lattice spectral functions then take the form
\begin{align}
 \rho_{S}(\om)  = &\, \frac{4\pi N_c}{N_s^3} \sum_\pv \delta\left(\om-2E_\pv\right), \\ 
 \rho_{P}(\om)  = &\, \frac{4\pi N_c}{N_s^3} \sum_\pv \hat\pv^2 \delta\left(\om-2E_\pv \right),
\end{align}
where the sums go over all momenta in the first Brillouin zone. Evaluating these numerically, as in Refs.\ \cite{Karsch:2003wy,Aarts:2005hg}, yields the spectral functions shown in Fig.\ \ref{fig:free}. Here $N_s$ is taken large enough to be in the spatial thermodynamic limit, while there is no dependence on $N_\tau$.
 Improving the dispersion relation will give better agreement between lattice and continuum results at small $\om$. The cusps result from reaching the edge of the Brillouin zone in the (1,0,0) or (1,1,0) direction ($+$ permutations). The maximal energy is determined by the maximal lattice momentum in the (1,1,1) direction, namely $\hat\pv^2=12$, and equals $a_\tau\om_{\rm max}^{\rm lat} = 0.503$ for the parameters used here.
 Comparing these results with those for free relativistic quarks \cite{Karsch:2003wy,Aarts:2005hg}, we conclude that the main difference is the temperature (or $N_\tau$) independence.

\end{document}